\documentclass[12pt]{article}
\usepackage{a4,amsmath,amssymb,cite,graphicx}
\def\Bbb{\mathbb}
\def\BZ{\Bbb Z} 
 
\def\BH{\mathbb{H}}

\catcode`@=11 \@addtoreset{equation}{section} \catcode`@=12

\begin{document}
\bibliographystyle{utphys}
\begin{titlepage}
\renewcommand{\thefootnote}{\fnsymbol{footnote}}
\noindent
{\tt IITM/PH/TH/2009/3}\hfill
{\tt arXiv:0907.1410v2}\\[4pt]
{\tt IMSc-2009-04-06} \hfill 
\hfill{\fbox{\textbf{v2.0; Dec 2009}}}

%\vspace{1.0cm}
\begin{center}
\large{\bf  BKM Lie superalgebras from
dyon spectra in $\BZ_N$-CHL orbifolds for composite $N$}
\end{center} 
\bigskip 
\begin{center}
Suresh Govindarajan\footnote{\texttt{suresh@physics.iitm.ac.in}} \\
\textit{Department of Physics, Indian Institute of Technology Madras,\\ Chennai 600036, INDIA.}
\\[5pt]
and \\[5pt] 
K. Gopala Krishna\footnote{\texttt{gkrishna@imsc.res.in}} \\
\textit{The Institute of Mathematical Sciences,\\
CIT Campus, Taramani, 
Chennai 600113 INDIA}
\end{center}

\begin{abstract}
We show that the generating function of electrically charged $\tfrac12$-BPS states in  $\mathcal{N}=4$ supersymmetric CHL $\mathbb{Z}_N$-orbifolds of the heterotic string on $T^6$ are given by multiplicative $\eta$-products. The $\eta$-products are determined by the cycle shape of the corresponding symplectic  involution in the dual type II  picture. This enables us to complete the construction of the genus-two Siegel modular forms due to David, Jatkar and Sen \texttt[arXiv:hep-th/0609109] for $\mathbb{Z}_N$-orbifolds when $N$ is non-prime. We study the $\mathbb{Z}_4$-CHL orbifold in detail and show that the associated Siegel modular forms, $\Phi_3(\mathbb{Z})$ and $\widetilde{\Phi}_3(\mathbb{Z})$, are given by the square of the product of three even genus-two  theta constants. Extending work by us as well as Cheng and Dabholkar, we show that the `square roots' of the two Siegel modular forms appear as the denominator formulae of two distinct  Borcherds-Kac-Moody (BKM) Lie superalgebras.  The BKM Lie superalgebra associated with the generating function of $\tfrac14$-BPS states, i.e., $\widetilde{\Phi}_3(\mathbb{Z})$ has a parabolic root system with a light-like Weyl vector and the walls of its fundamental Weyl chamber are mapped to the walls of marginal stability of the $\tfrac14$-BPS states.
\end{abstract}
\end{titlepage}
\setcounter{footnote}{0}
\section{Introduction}

More than a decade ago, Dijkgraaf, Verlinde and Verlinde (DVV) proposed a microscopic index formula for the degeneracy of $\tfrac14$-BPS dyons in heterotic string theory compactified on a six-torus\cite{Dijkgraaf:1996it}. Since then, and especially in the past few years, there has been a lot of progress in the microscopic counting of $\tfrac14$-BPS states and the ideas of DVV have been extended to a larger family of models in four-dimensional $\mathcal{N}=4$ compactifications in string theory. There has also emerged a promising new direction by studying the algebra of the $\tfrac14$-BPS states. The degeneracy of the $\tfrac14$-BPS states, in all the models, is given by a generating function which is a genus-two Siegel modular form whose weight and level depends on the model in consideration\cite{Jatkar:2005bh}. The `square roots' of these genus-two modular forms have been found to be related to a general class of infinite dimensional Lie algebras known as Borcherds-Kac-Moody (BKM) Lie superalgebras and this endows the degeneracy of the $\tfrac14$-BPS states with an underlying BKM Lie superalgebra structure\cite{Cheng:2008fc,Govindarajan:2008vi,Cheng:2008kt}. Following this insight, physical ideas of the theory such as the structure of the walls of marginal stability\cite{Sen:2007vb} have been understood from an algebraic point of view as the walls of the fundamental Weyl chamber\cite{Cheng:2008fc,Cheng:2008kt}.

In this work, we focus on the CHL orbifolds with $\mathcal{N}=4$ supersymmetry. These arise as a family of asymmetric $\mathbb{Z}_N$-orbifolds of the heterotic string compactified on $T^4 \times \widetilde{S}^1 \times S^1$\cite{Chaudhuri:1995dj}.  Sen and Jatkar, extending the work of DVV to the CHL orbifolds, constructed a family of genus-two Siegel modular forms $\widetilde{\Phi}_k(\mathbf{Z})$ that generate the degeneracy of the $\tfrac14$-BPS states analogous to the construction of DVV\cite{Jatkar:2005bh}. The weight, $k$, of the modular form is related to the orbifolding group $\BZ_N$ as $(k+2) = 24/(N+1)$ (when $N$ is prime and $(N+1)|24$). From these modular forms, the dyon degeneracy is given by a three-dimensional contour integral($\mathcal{C}$ is a suitably chosen contour\cite{Jatkar:2005bh}):
\begin{equation}
d(n,\ell,m)=64 \oint_{\mathcal{C}} d^3z\  \frac{e^{-2\pi i (nz_1+mz_3+\ell z_2)}}{\widetilde{\Phi}_k(\mathbf{Z})}\ ,
\end{equation} 
where $\mathbf{Z}\in \BH_2$, the Siegel upper-half space and $(n,\ell,m)=(\frac12\mathbf{q_e}^2, \mathbf{q_e}\cdot \mathbf{q_m}, \frac12\mathbf{q_m}^2)$ are the T-duality invariant combinations of electric and magnetic charges. 

An important application and non-trivial check for the veracity of the above degeneracy formula is to compute and compare the Bekenstein-Hawking-Wald entropy of blackholes (with torsion one) with the macroscopic derivation of the same, which it agrees with in the limit of large charges\cite{Jatkar:2005bh,Sen:2007vb}. The degeneracy formula also captures the change in degeneracy when one crosses a wall of marginal stability, through a subtle dependence of the contour integral on moduli, which is in keeping with the physical idea that certain dyonic configurations are not stable when one crosses over to a different region in the moduli space across a wall of marginal stability and this should reflect in the index\cite{Sen:2007vb}. Thus, the degeneracy formula captures important physical aspects  of dyon counting at a microscopic level.

Originally, Sen and Jatkar  considered the family of theories obtained by  $\BZ_N$ orbifolding  for the case of prime $N$ and obtained the modular forms that generate the degeneracy of $\tfrac14$-BPS states in these theories\cite{Jatkar:2005bh}. For the case of composite $N$, however, only the general behavior has been subsequently studied\cite{David:2006ud}.

The first example of composite $N$, occurs for $N = 4$, is an interesting as well as an important theory to understand. It has been predicted that the BKM Lie superalgebra structure underlying the degeneracy of the $\tfrac14$-BPS states in theories with $N>3$ do not exist\cite{Cheng:2008kt}. If this claim is true(we show that it is false), then the algebra of the $\tfrac14$-BPS states undergoes a transition between $N=3$ and $N=4$, and it would be educative to compare the $N=4$ theory with the $N = 1, 2, 3$ theories to understand just what sort of structure it is that generates this algebra for the dyonic degeneracies that is absent for the case of $N=4$. Before that, however, one needs to verify the claim explicitly by looking at the structure of the Siegel modular form that generates the degeneracy of $\tfrac14$-BPS states.

Parallel to the construction of ${\widetilde{\Phi}_k(\mathbf{Z})}$, encoding the degeneracy of the $\tfrac14$-BPS states, a related family of modular forms, denoted $\Phi_k(\mathbf{Z})$, that encode information on the $R^2$-corrections to the string effective action, were also constructed and studied\cite{Sen:2005pu}.These modular forms have also appeared recently in the work of 
Sen as twisted helicity trace indices in the $N=1$ theory\cite{Sen:2009md}. For the case of $N=1$ it turns out that $\widetilde{\Phi}_k(\mathbf{Z}) = \Phi_k(\mathbf{Z})$. For $N>1$, however, the two modular forms are different and one can also construct another family of  BKM Lie superalgebras associated with the family $\Phi_k(\mathbf{Z})$. Unlike in the case of ${\widetilde{\Phi}_k(\mathbf{Z})}$, the BKM structure for this family of modular forms does not undergo a radical change in structure as $N$ is changed and has been shown to exist for all prime $N$, even when $N>4$\cite{Govindarajan:2008vi}. In constructing the theory for $N = 4$, we also need to see if the BKM structure associated with the family  $\Phi_k(\mathbf{Z})$, that existed for the case of prime $N$, continues to exist for the case of composite $N$. We will show that this is indeed the case.

\noindent A summary of the main results of this paper is as follows:
\begin{enumerate}
\item We show that the generating function of $\tfrac12$-BPS states is given by multiplicative $\eta$-products. We obtain the $\eta$-products for \textit{all} groups that arise as symplectic involutions of $K3$.
\item The modular forms $\Phi_k(\mathbf{Z})$ and $\widetilde{\Phi}_k(\mathbf{Z})$ have been constructed for $N=4,6,8$ via the additive lift. Further, we have also worked out the systematics of the product formulae and given explicit expressions for $N=4$. It has also been shown that the $N=4$ modular forms can be written as a product of even genus-two theta constants.
\item The BKM Lie superalgebra for $\Delta_{3/2}(\mathbf{Z})$ (which we denote by $\mathcal{G}_4$( is shown to be similar to the ones appearing in \cite{Govindarajan:2008vi}. The Cartan matrix, Weyl vector and Weyl group remain unchanged by the orbifolding. However, the multiplicities of the imaginary simple roots do depend on the orbifolding. We provide evidence that the BKM Lie superalgebras $\mathcal{G}_N$ for N=2,3,4,5 are related to the dyons counted by the twisted helicity trace indices in heterotic string theory on $T^6$\cite{Sen:2009md}.
\item The BKM Lie superalgebra for $\widetilde{\Delta}_{3/2}(\mathbf{Z})$ is of parabolic type with infinite real simple roots (labelled by an integer) with Cartan matrix
\begin{equation*}
A^{(4)} = (a_{nm})\quad \textrm{where}\quad 
a_{nm}= 2 -4(n-m)^2\  , \eqno{\eqref{infiniteCartan}}
\end{equation*}
and a light-like Weyl vector. The walls of marginal stability for the $N=4$ model get mapped to the walls of the fundamental Weyl chamber of the BKM Lie superalgebra $\widetilde{\mathcal{G}}_4$
\item We also provide a proposal for modular forms for product groups as well as type II models. 
\end{enumerate}

The organization of the paper is as follows. In section 2, we discuss the relevant details of the model as well as provide some of the relevant details of the microscropic counting of dyon degeneracies as carried out in \cite{David:2006yn}. In section 3, we provide the construction of the modular forms $\Phi_k(\mathbf{Z})$ for $N=4,6,8$ via the additive lift. In section 3.1, we show that the generating function of degeneracy of $\tfrac12$-BPS states is given by a product of $\eta$-functions that satisfy a multiplicative property -- these are called $\eta$-products. This identification helps us to construct the weak Jacobi form that is the seed for the additive lift. In section 4, we construct product formulae for the modular forms for the $N=4$ orbifold ${\widetilde{\Phi}_3(\mathbf{Z})}$ and $\Phi_3(\mathbf{Z})$ -- the important details of the computation of twisted elliptic genera for $N=4,6,8$ are, however, relegated to appendix \ref{twistedellgen}.  In section 5, we show that the `square root' of the two modular forms (constructed in the previous section) do appear to  be the denominator formulae for two distinct BKM Lie superalgebras.  In section 5.2, we construct the BKM Lie superalgebra for the modular form $\widetilde{\Delta}_{3/2}(\mathbf{Z})$ and show that it has infinite real simple roots (reflecting the parabolic nature of the algebra) and the walls of the Weyl chamber do get mapped to walls of marginal stability of dyons. This is similar to what happened for $N=1,2,3$\cite{Cheng:2008fc,Cheng:2008kt} and is consistent with the general expectations for $N=4$ in \cite{Sen:2007vb}. In section 5.3,  we construct the BKM Lie superalgebra for the modular form $\Delta_{3/2}(\mathbf{Z})$ and show that it is consistent with the expectations in \cite{Govindarajan:2008vi}. In section 6, we discuss two natural generalizations of the models in this paper. First, we consider the case of product orbifolds of the form $\BZ_n\times \BZ_m$ and next we consider the case of $\mathcal{N}=4$ supersymmetric type II models that are orbifolds of type II string theory on a six-torus. We conclude in section 7 with a brief summary of our results. The appendices are used to provide technical results relevant for the paper. In particular, appendix   \ref{twistedellgen} determines the twisted elliptic genera using consistency conditions and reduces the computation to a few unknown constants that can be determined by the methods given in Sec. 4 for the $N=4$ orbifold.

\textbf{Note:} The modular form $\Delta_{3/2}(\mathbf{Z})$ has also been constructed by Gritsenko and Clery\cite{Gritsenko:2008}. Our results are in agreement with theirs even though our approaches differ.

\section{The model}

The model that we consider is a four-dimensional compactification of string theory with $\mathcal{N} =4$ supersymmetry. It has three perturbative formulations in terms of toroidally compactified heterotic string and type IIA/B string theory compactified on $K3 \times T^2$. We study the orbifolding of this theory by a $\BZ_N$ group such that the $\mathcal{N} =4$ supersymmetry of the unorbifolded theory is preserved. (See the review by Sen \cite{Sen:2007qy} and references therein for details.) 

Consider the four-dimensional heterotic string compactified on $T^4 \times \widetilde{S}^1 \times S^1$. The generator of the $\BZ_N$-orbifolding acts by a $1/N$ shift along the circle $S^1$ and a simultaneous $\BZ_N$ involution of the Narain lattice, $\Gamma^{20,4}$, of signature $(20,4)$ associated with the $T^4$. In the dual type IIA theory, the lattice $\Gamma^{20,4}$ is identified with $H^*(K3,\BZ)$ and the orbifolding group, $\BZ_N$, is a symplectic Nikulin involution combined with the $1/N$ shift of $S^1$. 

The massless spectrum of the four-dimensional $\mathcal{N} =4$ supersymmetric toroidally compactified heterotic string consists of one graviton multiplet together with $22$ vector multiplets. The massless spectrum depends on the orbifolding group. The orbifolding group acts trivially on the right-moving fermions, and all the $16$ supercharges are preserved even after the orbifolding. However, it acts non-trivially on the left-moving gauge degrees of freedom and hence the number of vector multiplets will be fewer. Also, the $1/N$ shift along the circle forces all the twisted sector states to be massive. Thus, the orbifolded theory has fewer massless vector multiplets down from the $22$ in the unorbifolded theory to $m=\textrm{rk}(\Gamma^\perp)-2$, where $\Gamma^\perp$ is the  sub-lattice of  $\Gamma^{20,4}$ that is invariant under the orbifold action.

The bosonic part of the low-energy effective action (up to two derivatives), in the variables of the heterotic description  is
\begin{multline}\label{lowenergy}
S=\int d^4x \sqrt{-g}\left[R - \frac{\partial_\mu\lambda \  \partial^\mu \bar{\lambda}}{2~ \textrm{Im}(\lambda)^2} +\frac18 \textrm{Tr}(\partial_\mu ML\ \partial^\mu M L) \right. \\ \left.-\frac14 \textrm{Im}(\lambda)\ F_{\mu\nu}LML\ F^{\mu\nu} +\frac14 \textrm{Re}(\lambda)\ F_{\mu\nu}L\ \widetilde{F}^{\mu\nu} \right]\ ,
\end{multline} 
where $L$ is a Lorentzian metric with signature $(6,m)$, $M$ is a 
$(6+m)\times (6+m)$ matrix valued scalar field satisfying $M^T=M$ 
and $M^TLM=L$ and $F_{\mu\nu}$ is a $(6+m)$ dimensional vector representing the  field 
strengths of the $(6+m)$ gauge fields.

The moduli space of the scalars is 
\begin{equation}
\big(\Gamma_1(N)\times SO(6,m;\BZ)\big)\bigg\backslash\! \left(\frac{SL(2)}{U(1)} \times \frac{SO(6,m)}{SO(6)\times SO(m)}\right). \end{equation}
$SO(6,m;\BZ)$ is the T-duality symmetry group and $\Gamma_1(N)\subset PSL(2,\BZ)$ given by 
\begin{equation}
\Gamma_1(N) = \left\{ \left(\begin{array}{cc}a & b \\ c & d \end{array}\right) \bigg| \  ad -bc =1,  \  c =0 \textrm{ mod }N, \quad a=d= 1 \textrm{ mod }N \right\}
\end{equation}
is the S-duality symmetry group that is manifest in the equations of motion and is compatible with the charge quantization. The fields that appear at low-energy can be organized into multiplets of these various symmetries. 
\begin{enumerate}
\item The heterotic dilaton combines with the axion (obtained by dualizing the antisymmetric tensor) to form the complex scalar $\lambda$ in the complex upper-half plane. 
\item The $(6+m)$ vector fields transform as an $SO(6,m;\mathbb{Z})$ vector under the T-duality group. Thus, the electric charges $\mathbf{q}_e$ (resp. magnetic charges $\mathbf{q}_m$) associated with these vector fields are also vectors (resp. co-vectors) of $SO(6,m,\mathbb{Z})$. Further, the electric and magnetic charges transform as a doublet under the $S$-duality group, $\Gamma_1(N)$.
\end{enumerate}
 Under the action of the S-duality group,  the charges of dyons and the complex scalar transform as follows
\begin{equation}
\lambda \rightarrow \frac{a \lambda + b}{c \lambda + d}\ , \quad \left( \begin{array}{c}\mathbf{q}_e \\ \mathbf{q}_m \end{array} \right) \rightarrow \left(\begin{array}{cc}a & b \\ c & d \end{array}\right) \left( \begin{array}{c}\mathbf{q}_e \\ \mathbf{q}_m \end{array} \right)\ .
\end{equation}
One can form three T-duality invariant scalars, $\mathbf{q}_e^2$, 
$\mathbf{q}_m^2$ and $\mathbf{q}_e\cdot\mathbf{q}_m$ from the 
charge vectors. These transform as a triplet of the S-duality 
group. Equivalently, we can write the triplet as a symmetric 
matrix:
\begin{equation}
\mathcal{Q}\equiv  \left(\begin{array}{cc}\mathbf{q}_e^2 
& \mathbf{q}_e\cdot\mathbf{q}_m \\ 
\mathbf{q}_e\cdot\mathbf{q}_m& \mathbf{q}_m^2\end{array}\right)\ .
\end{equation}
The $S$-duality transformation now is $\mathcal{Q}\rightarrow \gamma 
\cdot \mathcal{Q} \cdot \gamma^T$ with $\gamma\in \Gamma_1(N)$. The charges 
are quantized such that $N \mathbf{q}_e^2,\ \mathbf{q}_m^2 \in 2 
\BZ$ and $\mathbf{q}_e\cdot\mathbf{q}_m\in \BZ$. There exist many 
more invariants due to the discrete nature of  the T-duality 
group\cite{Banerjee:2007sr,Banerjee:2008ri} for $N=1$ and more appear when $N>1$. One such discrete invariant is the torsion $r=\textrm{gcd}(\mathbf{q}_e\wedge \mathbf{q}_m)$\cite{Dabholkar:2007vk}. In this paper, we will restrict our attention to $\tfrac14$-BPS dyons with torsion $r=1$.

There is also a parity transformation that enlarges the modular group from $PSL(2, \BZ)$ to $PGL(2, \BZ)$ and acts on the complex scalar and the dyons as\cite{Cheng:2008fc}  
\begin{equation}
R: \left( \begin{array}{c}\mathbf{q}_e \\ \mathbf{q}_m \end{array} \right) \rightarrow  \left( \begin{array}{c}\mathbf{q}_e \\ -\mathbf{q}_m \end{array} \right), \quad \lambda \rightarrow \bar{\lambda}\ .
\end{equation}
On adding the parity symmetry to  the S-duality group, $\Gamma_1(N)$, one obtains   the `extended S-duality symmetry group', $\widehat{\Gamma}_1(N)$\cite{Cheng:2008kt}.

\subsection{Microscopic counting of dyonic degeneracies}

In this subsection, we will discuss the microscopic counting of dyon degeneracies carried out by David and Sen\cite{David:2006yn,Sen:2007qy}. Consider type IIB string theory on $K3\times \widetilde{S}^1\times S^1$ modded by the $\mathbb{Z}_N$ symmetry that acts a symplectic involution of $K3$ and a $1/N$-shift of the $S^1$. The configuration considered in \cite{David:2006yn} consists of one D$5$-brane wrapping $K3\times S^1$, $Q_1$ D$1$-branes wrapped on $S^1$,  a single Kaluza-Klein(KK) monopole associated with the circle $\tilde{S}^1$ with negative magnetic charge, $-k$ units of momentum along $S^1$ and momentum $J$ along $\widetilde{S}^1$. This configuration corresponds to the BMPV black hole at the center of Taub-NUT space\cite{Breckenridge:1996sn}.The main idea used by David-Sen is to use the 4D-5D correspondence combined with known dualities to map the counting of states in this configuration to the counting of dyonic degeneracies in the  CHL string\cite{Gaiotto:2005gf}. 

Let $d(\mathbf{q}_e,\mathbf{q}_m)$ denote the number of bosonic minus fermionic quarter BPS supermultiplets carrying a given set of charges $(\mathbf{q}_e,\mathbf{q}_m)$ in the configuration described above. 
The dyonic charges of the configuration above are given by
\begin{equation}
\mathbf{q}_e^2=2k/N\ , \ \mathbf{q}_m^2= 2 (Q_1-1)\ , \ \mathbf{q}_e\cdot \mathbf{q}_m=J\ .
\end{equation}
The quantum numbers $k$ and $J$ can arise from three different sources:
\begin{enumerate}
\item The excitations of the Kaluza-Klein monopole carrying momentum $-l'_0/N$ along $S^1$.
\item The overall motion of the D$1$-D$5$ system in the background of the Kaluza-Klein monopole carrying momentum $-l_0/N$ along $S^1$ and $j_0$ along $\widetilde{S}^1$.
\item The motion of the $Q_1$ D$1$-branes in the worldvolume of the D5-brane carrying momentum $-L/N$ along $S^1$ and $J'$ along $\widetilde{S}^1$.
\end{enumerate}
Thus, we have 
\begin{equation}
 l'_{0} + l_0 + L = k\ , \quad j_0 + J' = J\ .
\end{equation}
In the weak coupling limit, one assumes that one can ignore the interaction between the three different sets of degrees of freedom and obtain the generating  function of dyonic degeneracies of the whole system as a product of the generating functions of each of the three separate pieces. Let $f(\rho,\sigma,v)$ denote the generating function of the whole system:
\begin{equation}
f(\rho,\sigma,v)= \sum_{k, Q_1,J}\  d(\mathbf{q}_e,\mathbf{q}_m)\ e^{2 \pi i \big[\sigma (Q_1-1)/N + \rho k + v J\big]}\ .
\end{equation}
Then, from the above argument it can be written as
\begin{multline}
f \left( \rho,\sigma,v \right) =\frac{1}{64}\  e^{-2\pi i\sigma/N}  \Big( \sum_{Q_1, L, J'} (-1)^{J'} d_{D1} (Q_1, L, J^{'})\ e^{2 \pi i (\sigma Q_1/N + \rho L + v J')} \Big) \\ 
\Big( \sum_{l_0, j_0} (-1)^{j_0} d_{CM} (l_0, j_0)\ e^{2 \pi i l_0 \rho + 2 \pi i j_0 v} \Big) \Big( \sum_{l'_{0}} d_{KK} (l'_{0})\ e^{2 \pi i l'_{0} \rho }\Big),
\end{multline}
where $d_{D1} (Q_1, L, J^{'})$ is the degeneracy of the $Q_1$ D$1$-branes moving in the plane of the D$5$-brane, $d_{CM}(l_0, j_0)$ is the degeneracy associated with the overall motion of the D$1$-D$5$ system in the background of the Kaluza-Klein monopole(i.e., its motion in Taub-NUT space), and $d_{KK} (l'_0)$ is the degeneracy associated with the excitations of the Kaluza-Klein monopole. The factor of $1/64$ removes the degeneracy of a single $\tfrac14$-BPS supermultiplet.

Let us write $f(\rho,\sigma,v)$ as
\begin{equation}
f(\rho,\sigma,v)=\big[\hat{\mathcal{E}}_{S^*(K3/\BZ_N)} 
(\rho,\sigma,v) 
\times \mathcal{E}_{\textrm{TN}}(\rho,v)\times  g(\rho) \big]^{-1}\ ,
\end{equation}
where
\begin{align*}
\big[\mathcal{E}_{S^*(K3/\BZ_N)} 
(\rho,\sigma,v)\big]^{-1} &\equiv  \sum_{Q_1, L, J'} (-1)^{J'} d_{D1} (Q_1, L, J^{'})\ e^{2 \pi i (\sigma Q_1/N + \rho L + v J')}\ , \\
\big[{\mathcal E}_{\textrm{TN}}(\rho,  v)\big]^{-1} &\equiv \frac1{4} \sum_{l_0, j_0} (-1)^{j_0} d_{CM} (l_0, j_0)\ e^{2 \pi i l_0 \rho + 2 \pi i j_0 v} \ ,\\
\big[g(\rho)\big]^{-1} &\equiv \frac1{16} \sum_{l'_{0}} d_{KK} (l'_{0})\ e^{2 \pi i l'_{0} \rho }\ .
\end{align*}

The computations in the appendices of David and Sen in \cite{David:2006yn} provide a microscopic understanding of the three different sources. We now summarize their results choosing a notation that is more or less identical to theirs.
\begin{enumerate}
\item The Taub-NUT space breaks eight of the sixteen supersymmetries in type IIB on $K3$ and quantization of its fermionic zero modes gives rise to a multiplicative factor of $16=2^{8/2}$. Following the chain of dualities, one sees that the Taub-NUT space gets mapped to the heterotic string wrapped on a $\BZ_N$-orbifold of the heterotic string. The degeneracy $d_{KK}(l'_0)$ corresponds to the degeneracy of the heterotic string in a twisted sector. Thus, $g(\rho/N)$ is the \textit{partition function} of the heterotic string (in a twisted sector) with the supersymmetric right-movers in their ground state. Thus, it can also be identified with the generating function of degeneracies of electrically charged $\frac12$-BPS states. We will provide an explicit expression for $g(\rho)$ for arbitrary $N$ in the next section as explicit formulae are known \textit{only} when $N$ is prime and divides $24$.
\item The D$1$-D$5$ system moving in $K3\times TN$ breaks four of the eight supersymmetries of type IIB string theory on $K3\times TN$. The quantization of its zero modes gives rise to a multiplicative factor of $4=2^{4/2}$. An explicit computation shows that 
\begin{equation}
\mathcal{E}_{\textrm{TN}}(\rho,v)=\left[\frac{\vartheta_1(N\rho,v)}{\eta(N\rho)^3}\right]^2 \ . 
\end{equation}
\item $\mathcal{E}_{S^*(K3/\BZ_N)} 
(\rho,\sigma,v)$ is the second-quantized elliptic genus of $K3/\BZ_N$\cite{Dijkgraaf:1996xw}.
\end{enumerate}
David and Sen further show that 
\begin{equation}
f(\rho,\sigma,v)=\frac1{\widetilde{\Phi}_k(\rho, \sigma, v)}=\frac1{\widetilde{\Phi}_k(\sigma/N, N\rho, v)}\ ,
\end{equation}
in the process obtaining a product representation for the generating function of $\tfrac14$-BPS states, $\widetilde{\Phi}_k(\mathbf{Z})$. Further, we note that the product 
$$
\phi(\sigma,v)=\mathcal{E}_{\textrm{TN}}(\sigma/N,  v)\times  g(\sigma)\ ,
$$ 
is the Jacobi form that is the additive seed for the modular form $\Phi_k(\mathbf{Z})$\cite{Jatkar:2005bh}. This enables us to determine this modular form once we explicitly determine $g(\rho)$ in the next section.

\section{The additive lift}

In the section, we first obtain  the generating function, $g(\rho)$, of the degeneracy of electrically charged $\tfrac12$-BPS states. This is used to obtain the Jacobi form that generates the additive lift and finally we construct the corresponding modular form.

\subsection{Counting $\frac12$-BPS states}

The counting of the degeneracy of $\tfrac12$-BPS states of a given electric charge is mapped to the counting of states of the heterotic string with the right-movers\footnote{In our convention, right movers are taken to be supersymmetric and left movers are bosonic in the heterotic string.} in the ground state\cite{Sen:2005ch,Dabholkar:2005dt,Dabholkar:2005by}. While this is conceptually easy to compute, for orbifolds, the contributions from the different sectors to the degeneracy need to be added up. This computation has been carried out by Sen who showed that, up to exponentially suppressed terms (for large charges), the leading contribution arises from the twisted sectors and the asymptotic expansion takes a simple form (given in Eq. \eqref{asympg} below)\cite{Sen:2005ch}. In this subsection, we show that this asymptotic expansion is consistent with a product of $\eta$-functions that we shall call $\eta$-products. This result, in  a sense, is a simplification and extension of the analysis of Sen\cite{Sen:2005ch}.

\subsubsection{Heterotic string on $T^6$}

(Electric) $\tfrac12$-BPS excitations of the heterotic string carrying charge $N\equiv \tfrac12 \mathbf{q}_e^2$ are obtained by choosing the supersymmetric (right-moving) sector to be in the ground state. The level matching condition becomes
\begin{equation}
%\label{ }
-\frac12 \mathbf{q}_e^2 + N_L =1\ , 
\end{equation}
where $\mathbf{q}_e\in \Gamma^{22,6}$ and $N_L$ is the oscillator contribution to $L_0$ in the bosonic (left-moving) sector. Thus, we see that 
$$
n= \tfrac12 \mathbf{q}_e^2=N_L-1\ .
$$
Let $d(n)$ represent the number of configurations of the heterotic string with electric charge such that $ \tfrac12 \mathbf{q}_e^2=n$. The level matching condition implies that we need to count the number of states with total oscillator number $N_L=(n+1)$. The generating function for these states  is
\begin{equation}
%\label{ }
\frac{16}{\eta(\tau)^{24}} = \sum_{n=-1}^\infty d(n) \ q^n\ ,
\end{equation}
where the factor of $16$ accounts for the degeneracy of a $\tfrac12$-BPS multiplet -- this is the degeneracy of the Ramond ground state in the right-moving sector.

\subsubsection{The CHL  orbifold of the heterotic string on $T^6$}

In the CHL orbifold, the electric charge takes values in a lattice $\Gamma^\perp\subset \Gamma^{22,6}$  of signature $(22-2\hat{k},6)=(2k+2,6)$ that is not self-dual.\footnote{An explicit description of the relevant lattices may be obtained by combining the work of Garbagnati and Sarti\cite{Garbagnati:2008} (who work out the invariant lattices under all abelian symplectic involutions of $K3$) and six-dimensional string-string duality that relates the type II string compactified on $K3$ to the heterotic string compactified on $T^4$.} Here $\Gamma^\perp$ is the sub-lattice of $\Gamma^{22,6}$ that is invariant under the action of the orbifold group. Let $\textrm{vol}^\perp$ be the volume of the unit cell in $\Gamma^\perp$. Define the generating function of the degeneracies $d(n)$ of $\tfrac12$-BPS states as follows:
\begin{equation}
%\label{ }
\frac{16}{g_\rho(\tau/N)}\equiv \sum_{n=-1}^\infty d(n) \ q^{n/N}\ ,
\end{equation}
for the $\mathbb{Z}_N$-CHL orbifold taking into account that the electric charge is quantized such that $N \mathbf{q}_e^2\in 2\mathbb{Z}$. Setting $\tau=i\mu/2\pi$, Sen has shown that in the limit $\mu\rightarrow 0$, one has\cite{Sen:2005ch}
\begin{equation}
\label{asympg}
\lim_{\mu\rightarrow0} \frac1{g_\rho(i\mu/2\pi N)}=16\  e^{4\pi^2/\mu}\ \left(\frac{\mu}{2\pi}\right)^{(k+2)/2}\  (\textrm{vol}^\perp)^{1/2} + \cdots
\end{equation}
where the ellipsis indicate exponentially suppressed terms. We will make an ansatz for $g_\rho(\tau)$ in the form of an $\eta$-product\footnote{The ansatz is based on the observation that this $\eta$-product is the modular transform of the the oscillator contribution to the $g$-twisted partition function: $\underset{\ ~g}{1\ \framebox[10pt]{\phantom{a}}}$. Not unsurprisingly, the same function appears in the additive seed for the modular form associated with the twisted index of Sen\cite{Sen:2009md}.}
\begin{equation}
%\label{ }
g_\rho(\tau)=\prod_{r=1}^N \eta(r\tau)^{a_r}=\eta(\tau)^{a_1}\eta(2\tau)^{a_2}\cdots \eta(N\tau)^{a_N}\ .
\end{equation}
We identify the above $\eta$-product with the `cycle shape' $\rho=1^{a_1}2^{a_2}\cdots N^{a_N}$. The $\eta$-product has to satisfy the following conditions:
\begin{enumerate}
\item The asymptotic behaviour of $g_\rho(\tau)$ given in Eq. \eqref{asympg} requires
\begin{align}
\label{compatcondn}
  \left(N a_1 +N \tfrac{a_2}2 +\cdots +a_N \right) &=24  \ ,\nonumber \\
   a_1+a_2+\cdots+ a_N&=2(k+2)   \ , \\
   \big(1^{a_1}2^{a_2}\cdots N^{a_N}\big)^{-1}&=\textrm{vol}^\perp\ . \nonumber
\end{align}
The last condition involving the volume of the unit cell is exactly what one expects for an orbifold action on the basis vectors of the self-dual lattice $\Gamma^{20,4}\subset \Gamma^{22,4}$ corresponding to the cycle shape $\rho$.
\item Considering $\BZ_N$ as a cyclic permutation, one sees that the only permitted cycles are of length $r$ such that $r|N$. One therefore imposes
 $a_r=0$ unless $r|N$. Thus, when $N$ is prime, only $a_1$ and $a_N$ are non-zero which agrees with known results.
\item
We will show later that the cycle shapes associated with Nikulin involutions satisfy an additional property -- they are \textit{balanced}. This  implies that $a_1=a_{N}$ among other things. It also implies that the first equation in Eq. \eqref{compatcondn} can be rewritten as
\begin{equation}
a_1 + 2 a_2 +\cdots + N a_N = 24\ .
\end{equation}
\end{enumerate}
 These conditions \textit{uniquely} fix the form of $g_\rho(\tau)$. When $N$ is prime, one sees that $a_1=a_N=\tfrac{24}{N+1}$ in agreement with known results\cite{Jatkar:2005bh}.

\subsubsection{Symplectic automorphisms of $K3$ and the Mathieu group $\mathbf{M_{24}}$}

In this subsection, we will consider the dual description of the CHL orbifold as a supersymmetric orbifold of type II string theory on $K3\times T^2$. This will provide an understanding of the cycle shapes that appear in the $\tfrac12$-BPS state counting. The orbifold group acts on the $K3$ as a symplectic (Nikulin) involution -- it acts trivially on  the nowhere vanishing $(2,0)$ holomorphic form. Mukai showed that any finite group of symplectic automorphisms of a $K3$ surface is a subgroup of the  Mathieu group, $M_{23}$\cite{Mukai:1988}.

To better understand Mukai's result,
consider a symplectic automorphism of $K3$, $\sigma$, of finite order $n$ (it is known that $n\leq8$). He observed that the number of fixed points, $\varepsilon(n)$ (which depends only on 
the order of  $\sigma$) is given by
\begin{equation*}
%\label{fixedpointformula}
\varepsilon(n)=\frac{24}{n\prod_{p|n}(1+\tfrac1p)}\ ,
\end{equation*}
and happens to match the number of fixed points for a similar element of the Mathieu group, $M_{23}$. The Mathieu group $M_{24}$ can be represented  as a permutation group acting on a set with $24$ elements. Then, $M_{23}$ is the subgroup of $M_{24}$ that preserves one element of the set.
Mukai then showed that if $G$ is a finite group of  symplectic automorphisms of $K3$, then
\begin{itemize}
\item[(i)] $G$ acts as a permutation on the generators of $H^*(K3,\mathbb{Z})$ and can be embedded as a  subgroup of $M_{23}$. 
\item[(ii)] $G$ necessarily has at least five fixed points, one arising from $H^{0,0}(K3)$, $H^{2,0}(K3)$, $H^{1,1}(K3)$, $H^{0,2}(K3)$ and $H^{2,2}(K3)$. The only non-trivial part is that there is at least one fixed point in $H^{1,1}(K3)$. 
\end{itemize}

The embedding of $G$ into $M_{23}\subset M_{24}$ enables one to use known properties of $M_{24}$. In particular, Conway and Norton have shown that \textit{any} element of $M_{24}$ has a \textit{balanced} cycle shape\cite{Conway:1979}. Recall that any permutation (of order $n$) may be represented by its cycle shape: 
\begin{equation}
\label{cycleshape}
\rho\equiv 1^{a_1}2^{a_2}\cdots n^{a_n}\ .
\end{equation}
 A cycle shape, $\rho$, is said to be balanced if there exists a positive integer $M$ such that $\big(\tfrac{M}1\big)^{a_1} \big(\tfrac{M}2\big)^{a_2}\cdots \big(\tfrac{M}n\big)^{a_n}$ is the same as $\rho$. Since dim$(H^*(K3))=24$, one also has the condition \begin{equation}
\label{cyclecondition}
\sum_i i~a_i=24\ .
\end{equation}
As an example, the cycle shape $1^4 2^2 4^4$ is  balanced  with $M=4$ and satisfies the above condition. Now given a balanced cycle shape, $\rho$, consider the function $g_\rho(\tau)$ defined by the following product of $\eta$-functions:
\begin{equation}
%\label{ }
\rho \longmapsto g_\rho(\tau)\equiv \eta(\tau)^{a_1}\eta(2\tau)^{a_2} \cdots  \eta(n\tau)^{a_n}\ .
\end{equation}
Note that when the condition \eqref{cyclecondition} is satisfied, $g_{\rho}(\tau)$ has no fractional exponents in its Fourier expansion about the cusp at infinity. One has
\begin{equation}
\label{ }
g_{\rho}(\tau) =\sum_{m=1}^\infty a_m\ q^m \ ,\textrm{ with } a_1=1\ ,
\end{equation}
where $q=\exp(2\pi i \tau)$. Dummit, Kisilevsky and McKay\cite{Dummit:1985} (see also \cite{Mason:1985}) considered such functions after imposing an additional property called \textit{multiplicativity}. A function $g(\tau)=\sum_n a_n q^n$ is multiplicative\footnote{Martin imposes a more stringent condition\cite{MR1376550} -- he requires that the function and its image under the Fricke involution must be Hecke eigenforms. It turns out that all the examples that we consider satisfy the stronger condition.} if $a_{nm}=a_n a_m$ when gcd$(n,m)=1$. 

By means of a computer search among the $1575$ partitions of $24$ (this is equivalent to all  solutions of Eq.\eqref{cyclecondition}), Dummit et. al. found a set of thirty multiplicative $\eta$-products each associated with a cycle that was balanced. In Table \ref{Dummit}, we reproduce their table restricting to shapes with $M\leq 16$ after adding a couple of columns that are relevant to this paper. The last column is the discrete group $G$ that is an automorphism of $K3$ which corresponds to the cycle shape $\rho$ -- this has been added by us. The groups have been identified by extracting the cycle shape from the discussion in Chaudhuri and Lowe\cite{Chaudhuri:1995dj} (see also proposition 5.1 in \cite{Garbagnati:2008}). It is interesting to note that \textit{all} cycle shapes that appear in Table \ref{Dummit} arise from the action of Nikulin involutions on $K3$ -- this includes product groups such as $\mathbb{Z}_2\times \mathbb{Z}_2$. In examples involving product groups, the $\eta$-products are actually of level $N<M$ and we have indicated the true level $N$ in a separate column.
\begin{table}[t]
 \newcommand\T{\rule{0pt}{2.6ex}}
 \centering \begin{tabular}{|c|c|c|c|c|c|}\hline
   {\small \bf Cycle shape} $\rho$ &  $ (k+2) $ &$\chi\!\left(\begin{smallmatrix}a&b\\ c & d \end{smallmatrix}\right)$\T & $M$ &$N$& $G$ \\[3pt] \hline
 $1^{24}$  & $ 12 $&  & $ 1 $ & $ 1 $& \\ \hline
 $1^82^8 $ & $ 8 $&  & $ 2 $ & $ 2 $  &$ \mathbb{Z}_2$ \\ \hline
 $1^63^6$ & $ 6 $&  & $ 3 $& $ 3 $&$ \mathbb{Z}_3$ \\ \hline
 $2^{12}$ & $ 6 $ &  &  $4$ &$2 $ & $ \mathbb{Z}_2\times \mathbb{Z}_2 $ \\ \hline
 $1^42^24^4$ & $5$ & $ \left(\tfrac{-1}{~d}\right)\T $ & $4$ & $4$& $ \mathbb{Z}_4$ \\[3pt] \hline
  $1^45^4$ &$4$ &  & $5$ & $5$&$ \mathbb{Z}_5$ \\ \hline
 $1^22^23^2 6^2$ & $4$ &  & $6$ & $6$ &$ \mathbb{Z}_6$ \\ \hline
 $2^44^4$ & $4$&  &$ 8 $& $4$ & $ \mathbb{Z}_2\times \mathbb{Z}_4 $ \\ \hline
 $ 3^8$ &  $4$ & &$ 9 $ & $3$ & $ \mathbb{Z}_3\times \mathbb{Z}_3 $ \\ \hline
 $1^37^3$&  $3$& $ \left(\tfrac{-7}{~d}\right)\T $  & $7$ & $7$&$ \mathbb{Z}_7$ \\[3pt] \hline
 $1^22^1 4^1 8^2$ & $3$ & $ \left(\tfrac{-2}{~d}\right)\T $  & $8$ & $8$  &$ \mathbb{Z}_8$ \\[3pt] \hline
 $ 2^3 6^3$ & $3$ & $ \left(\tfrac{-3}{~d}\right)\T $ & $12$ &$6$ & $ \mathbb{Z}_2\times \mathbb{Z}_6 $ \\[3pt] \hline
 $ 4^6 $ & $3$ & $ \left(\tfrac{-1}{~d}\right) \T $ &  $ 16 $ & $4 $ &$ \mathbb{Z}_4\times \mathbb{Z}_4 $ \\[3pt] \hline
 $1^2 11^2$ &  $ 2 $ & & $ 11 $& $ 11 $ &$ \mathbb{Z}_{11}$ \\ \hline
\end{tabular}
  \caption{The function $g_\rho(\tau)$ is a modular form of weight $(k+2)$, generalized level $M$ (true level $N$) and  character $\chi$. Only non-trivial characters are indicated in column 3. }\label{Dummit}
\end{table}

\subsection{The additive lift}

Consider the weak Jacobi form of weight $k$, index $1$ and level $N$
\begin{equation}
\label{additiveseed}
\phi_{k,1}(z_1,z_2) = \frac{\vartheta_1(z_1,z_2)^2}{\eta(z_1)^6} \ g_\rho(z_1)=
\sum_{n,\ell} a(n,\ell)\ q^n r^\ell \ .
\end{equation}
We conjecture that this Jacobi form is the seed for the additive (Maa\ss) lift leading to the genus-two Siegel modular form $\Phi_k(\mathbf{Z})$ when $G=\mathbb{Z}_N$.
Note that when $N$ is prime and $(N+1)$ divides $24$, then this agrees with the additive seed given in \cite{Jatkar:2005bh} as the cycle shape is $1^{k+2}N^{k+2}$ as given in the Table \ref{Dummit}. When $N$ is composite, the cycle shape is as given in Table \ref{Dummit}. The formula for $\Phi_k(\mathbf{Z})$ is given by the Fourier coefficients, $a(n,\ell)$,  of the additive seed 
\begin{equation}
%\label{ }
\Phi_k(\mathbf{Z})\equiv \sum_{(n,\ell,m)>0}\ \  
\sum_{d | (n,\ell,m)} \ \chi(d)\  
d^{k-1}\ a\left(\tfrac{nm}{d^2},\tfrac\ell{d}\right)\ 
q^{n} r^{\ell} s^{m}\ ,
\end{equation}
where $$(n,\ell,m)>0 \textrm{ implies  }n,m\in \BZ_+\ , \ \ell\in\BZ \textrm{ and }
(4nm-\ell^2)>0\ .$$
 In the above formula, the weight $k$ and  the character $\chi$ are as given in Table \ref{Dummit}. 
 
 As discussed by Jatkar and Sen\cite{Jatkar:2005bh}, the generating function of dyonic degeneracies, $\widetilde{\Phi}_k(\mathbf{Z})$, is given by expansion of the modular form, $ \Phi_k(\mathbf{Z})$, about another inequivalent cusp. Let
 \begin{equation}\label{tildedef}
 \widetilde{\Phi}_k(\mathbf{Z}) \equiv (\textrm{vol}^\perp)^{1/2} \ z_1^{-k} \
\Phi_k(\mathbf{\widetilde{Z}})\ ,
\end{equation}
with
\[
\tilde{z}_1 = -1/z_1\quad,\quad \tilde{z}_2 = z_2/z_1\quad, \quad 
\tilde{z}_3 = z_3 -z_2^2/z_1\ .
\]
We have chosen a  normalization for $\widetilde{\Phi}_k(\mathbf{Z})$ that differs from the one used in \cite{Jatkar:2005bh} but agrees with the one used in \cite{David:2006ud}.
Consider $\tfrac14$-BPS dyons with charges $\mathbf{q}_e$ and $\mathbf{q}_m$ such that $2 n =N \mathbf{q}_e^2$, 
$2 m=\mathbf{q}_m^2$ and $\ell=\mathbf{q}_e\cdot\mathbf{q}_m$. Then, the degeneracy $d(n,\ell,m)$ of dyons with these charges is generated by
\begin{equation}
 \frac{64}{\widetilde{\Phi}_k(\mathbf{Z})}= \sum_{n,\ell,m} d(n,\ell,m)\ q^{n/N} r^{\ell} s^{m}\ .
\end{equation}
A similar additive lift for $\widetilde{\Phi}_k(\mathbf{Z})$ is given by the following seed:
\begin{equation}
\label{tildeadditiveseed}
\widetilde{\phi}_{k,1}(z_1,z_2) = \frac{\vartheta_1(z_1,z_2)^2}{\eta(z_1)^6} \ g_\rho(z_1/N)\ .
\end{equation}

We now provide detailed expressions for the genus-two modular forms $\Phi_k(\mathbf{Z})$ for the $\mathbb{Z}_N$-CHL  orbifolds for $N=4,6,8$. 

\subsubsection{$N=4$}

From Table \ref{Dummit}, we see that $k=3$ for $N=4$. The seed for the additive lift is
\begin{equation}
%\label{ }
\phi_{3,1}(z_1,z_2) = \frac{\vartheta_1(z_1,z_2)^2}{\eta(z_1)^2} \ \eta(2z_1)^2 \eta(4z_1)^4=
\sum_{n,\ell} a(n,\ell)\ q^n r^\ell \ .
\end{equation}
The additive lift is ($a(n,\ell)$ is as defined by the above equation)
\begin{equation}
%\label{ }
\Phi_3(\mathbf{Z})\equiv \sum_{(n,\ell,m)>0}\ \  
\sum_{d | (n,\ell,m)} \  \left(\tfrac{-1}{~d}\right) 
d^{k-1}\ a\left(\tfrac{nm}{d^2},\tfrac\ell{d}\right)\ 
q^{n} r^{\ell} s^{m}\ ,
\end{equation}
where the Jacobi symbol $ \left(\tfrac{-1}{~d}\right)$ is $+1$ when $d=1\textrm{ mod }4$;  $-1$ when $d=3\textrm{ mod }4$ and $0$ otherwise. This is a Siegel modular form at level four and character $\psi_4(\gamma)$ where
\begin{equation}
\psi_4(\gamma)=\left(\frac{-1}{\det D}\right)\ \textrm{for } \gamma=\begin{pmatrix}A & B \\ C & D\end{pmatrix} \in G_0(4)\ ,
\end{equation}
where $G_0(4)$ is the level four subgroup of $Sp(2,\BZ)$\cite{Aoki:2005}.

\subsubsection{$N=6$}

From Table \ref{Dummit}, we see that $k=2$ for $N=6$. The seed for the additive lift is
\begin{equation}
%\label{ }
\phi_{2,1}(z_1,z_2) = \frac{\vartheta_1(z_1,z_2)^2}{\eta(z_1)^4} \ \eta(2z_1)^2 \eta(3z_1)^2\eta(6z_1)^2=
\sum_{n,\ell} a(n,\ell)\ q^n r^\ell \ .
\end{equation}
The additive lift is then ($a(n,\ell)$ is as defined by the above equation)
\begin{equation}
%\label{ }
\Phi_2(\mathbf{Z})\equiv \sum_{(n,\ell,m)>0}\ \  
\sum_{\substack{d | (n,\ell,m)\\ d=1,5\textrm{ mod }6}} \  
d^{k-1}\ a\left(\tfrac{nm}{d^2},\tfrac\ell{d}\right)\ 
q^{n} r^{\ell} s^{m}\ .
\end{equation}

\subsubsection{$N=8$}

From Table \ref{Dummit}, we see that $k=1$ for $N=8$. The seed for the additive lift is
\begin{equation}
%\label{ }
\phi_{1,1}(z_1,z_2) = \frac{\vartheta_1(z_1,z_2)^2}{\eta(z_1)^4} \ \eta(2z_1) \eta(4z_1)\eta(8z_1)^2=
\sum_{n,\ell} a(n,\ell)\ q^n r^\ell \ .
\end{equation}
The additive lift is then ($a(n,\ell)$ is as defined by the above equation)
\begin{equation}
%\label{ }
\Phi_1(\mathbf{Z})\equiv \sum_{(n,\ell,m)>0}\ \  
\sum_{d | (n,\ell,m)} \  \left(\tfrac{-2}{~d}\right) 
d^{k-1}\ a\left(\tfrac{nm}{d^2},\tfrac\ell{d}\right)\ 
q^{n} r^{\ell} s^{m}\ ,
\end{equation}
where the Jacobi symbol $ \left(\tfrac{-2}{~d}\right)$ is $+1$ when $d=1,3\textrm{ mod }8$; $-1$ when $d=5,7\textrm{ mod }8$ and $0$ otherwise. This is a Siegel modular form at level eight and character $\left(\tfrac{-2}{\det D}\right)$.

\section{Product formulae}

The product formulae for $\Phi_k(\mathbf{Z})$ as well as $\widetilde{\Phi}_k(\mathbf{Z})$ are given in terms of the coefficients of the Fourier expansion of the twisted elliptic genera\cite{David:2006ji}. The twisted elliptic genus for a $ \mathbb{Z}_N $-orbifold of $K3$ is defined as:
\begin{equation}
F^{a,b}(\tau,z) = \frac1N \textrm{Tr}_{RR,g^a} \Big((-)^{F_L+F_R} g^b q^{L_0}\bar{q}^{\bar{L}_0} e^{2\pi\imath z F_L}\Big)\ ,\quad 0\leq a\leq (N-1)\ ,
\end{equation}
where $ g $ generates $ \mathbb{Z}_N $ and $ q=\exp(2\pi \imath \tau) $. The twisted elliptic genera are weak Jacobi forms of weight zero, index one and level $N$\cite{David:2006ji}.   The Fourier expansion of the Jacobi form are 
\begin{equation}
F^{a,b}(\tau,z)= \sum_{m=0}^1 \sum_{\ell\in 2\mathbb{Z}+m,n\in \mathbb{Z}/N}c_m^{a,b}(4n-\ell^2)\ q^n r^\ell\ , \label{fourierdefs}
\end{equation}
where  $r=\exp(2\pi i z)$. We will also write $c^{a,b}(n,\ell)$ for the Fourier coefficient $c_m^{a,b}(4n-\ell^2)$.

In appendix \ref{twistedellgen}, we determine the twisted elliptic genera using consistency conditions based on their modular properties. When $N$ is prime, these conditions uniquely fix the twisted elliptic genera. For composite $N$, there remain undetermined parameters. These parameters are fixed by requiring that the product formula is compatible with the product form of the additive seed given in Eq. \eqref{additiveseed}. We illustrate the procedure for $N=4$.

\subsection{Product formula for $ \Phi_3(\mathbf{Z}) $}
Define 
\begin{equation}
\widehat{F}^{a}(\tau,z) =\sum_{b=0}^{3} F^{a,b}(\tau,z)\ ,
\end{equation}
and let $\hat{c}^a(n,\ell)$ be its Fourier coefficients. The product form given by David, Jatkar and Sen can be rewritten as\cite{David:2006ji}
\begin{equation}
%\boxed{
\Phi_3(\mathbf{Z})= q r s\!\! \prod_{(n,\ell,m)}\! \Big(1-q^nr^\ell
s^m\Big)^{\hat{c}^0-\hat{c}^2}\!\!\times
 \Big(1-\big(q^nr^\ell s^m\big)^2\Big)^{\hat{c}^2-\hat{c}^1}\!\!\times
 \Big(1-\big(q^nr^\ell s^m\big)^4\Big)^{\hat{c}^1}
%}
\end{equation}
where we have not written out the argument of $\hat{c}^a$ -- it is $(nm,\ell)$ in all occurrences above to reduce the length of the equation.  Other methods of generating product formulae have been used in \cite{Aoki:2005,Dabholkar:2006xa,Govindarajan:2008vi}.

\noindent Specializing the general formulae in appendix \ref{twistedellgen} to the case of $N=4$, we obtain
\begin{eqnarray}
\widehat{F}^{0}(\tau,z) &=& \tfrac{10}3 A(\tau,z) + (2b+\tfrac13)E_2(\tau)B(\tau,z) + (\tfrac56-2b)E_4(\tau)B(\tau,z)\nonumber \\
\widehat{F}^{1}(\tau,z) &=&\tfrac{4}3 A(\tau,z) -2b E_2(\tau)B(\tau,z) - (\tfrac5{12}-b)E_4(\tau)B(\tau,z)\\
\widehat{F}^{2}(\tau,z) &=& 2 A(\tau,z) + \tfrac12 E_2(\tau)B(\tau,z) - (\tfrac56-2b)E_4(\tau)B(\tau,z) \nonumber \ ,
\end{eqnarray}
where $A(\tau,z)$ and $B(\tau,z)$ are as defined in Eq. \eqref{ABdef}. This
leads to  formulae for the first two Fourier coefficients:
\begin{eqnarray}
\hat{c}^0(-1)=\tfrac56 +\tfrac13+\tfrac56=2 &\  ,\  & \hat{c}^0(0)=\tfrac{25}3-\tfrac73=6\ ,  \nonumber \\
\hat{c}^1(-1)=\tfrac13 -\tfrac5{12}-b=-b-\tfrac1{12} &\  ,\  & \hat{c}^1(0)=\tfrac{25}6+2b\ ,\\
\hat{c}^2(-1)=\tfrac12 +\tfrac12-\tfrac56+2b=2b+\tfrac16 &\  ,\ & \hat{c}^2(0)=\tfrac{17}3-4b \ .\nonumber
\end{eqnarray}
We need $\hat{c}^1(-1)=\hat{c}^2(-1)=0$ else we will have terms of the type $(1-r^2)$ and $(1-r^4)$ in the product expansion for $\Phi_3(\mathbf{Z})$. This fixes the unfixed constant $b=-1/12$. We can now write out all the terms with $m=0$ in the product formulae as we now have determined that $\hat{c}^1(0)=4$ and $\hat{c}^2(0)=6$. These give rise to terms of the form 
\begin{equation*}
\prod_{n=1}^\infty (1-q^n)^0 (1-q^{2n})^2 (1-q^{4n})^4\ .
\end{equation*}
This agrees with the (infinite set of) terms that appear from the product expansion of the additive seed:
\begin{equation*}
\phi_{3,1}(\tau,z)=\frac{\vartheta_1^2(\tau,z)}{\eta(\tau)^6} \eta(\tau)^4 \eta(2\tau)^2 \eta(4\tau)^4\ .
\end{equation*}
Since we have fixed the constant $b$, we can now write exact expressions for the $F^{a,b}(\tau,z)$:
\begin{eqnarray}
F^{0,0}(\tau,z)&=&2 A(\tau,z) \nonumber \\
F^{0,1}(\tau,z)&=& F^{0,3}(\tau,z)=\tfrac13 A(\tau,z)+\Big[ -\tfrac1{12} E_2(\tau)+\tfrac12 E_4(\tau)\Big] B(\tau,z)\nonumber  \\
F^{0,2}(\tau,z)&=& \tfrac23 A(\tau,z)+\tfrac13 E_2(\tau)B(\tau,z)\\
F^{1,k}(\tau,z)&=&F^{3,3k}(\tau,z)=\tfrac13 A(\tau,z)+\Big[ -\tfrac1{24} E_2\big(\tfrac{\tau+k}2\big)+\tfrac18 E_4\big(\tfrac{\tau+k}4\big)\Big] B(\tau,z)\nonumber  \\
F^{2,2k}(\tau,z)&=& \tfrac23 A(\tau,z)-\tfrac1{6} E_2\big(\tfrac{\tau+k}2\big)B(\tau,z) \nonumber \\
F^{2,2k+1}(\tau,z)&=& \tfrac13 A(\tau,z)+\Big[ \tfrac5{12} E_2(\tau)-\tfrac12 E_4(\tau)\Big] B(\tau,z)\nonumber
\end{eqnarray}
and
\begin{eqnarray}
\widehat{F}^{0}(\tau,z) &=& \tfrac{10}3 A(\tau,z) + \tfrac16E_2(\tau)B(\tau,z) + E_4(\tau)B(\tau,z) \ ,\nonumber \\
\widehat{F}^{1}(\tau,z) &=&\tfrac{4}3 A(\tau,z) +\tfrac16 E_2(\tau)B(\tau,z) - \tfrac12 E_4(\tau)B(\tau,z)\ ,\\
\widehat{F}^{2}(\tau,z) &=& 2 A(\tau,z) + \tfrac12 E_2(\tau)B(\tau,z) - E_4(\tau)B(\tau,z)
\nonumber\ .
\end{eqnarray}
Note that $ \sum_{r=0}^3\widehat{F}^{r}(\tau,z)=8 A(\tau,z) + (E_2(\tau) -E_4(\tau)) B(\tau,z)$. This appears to \textit{disagree} with the observation in David, Jatkar and Sen\cite{David:2006ru} that the sum should give  the elliptic genus of $K3$. Their prediction is that  $\sum_{r=0}^3\widehat{F}^{r}(\tau,z)=8 A(\tau,z)$. The other terms proportional to $B(\tau,z)$ are expected to vanish. However, the terms are such that the first two Fourier coefficients vanish and do not conflict with geometrical quantities of $K3$. So it agrees with their observation in a weaker sense.

We have been able to show that $ \Phi_3(\mathbf{Z}) $  can be written as the square of the product of three even genus-two theta constants. One has
\begin{equation}
\Phi_3(\mathbf{Z})=\left(\frac18\ \theta\!\left[\begin{smallmatrix} 1\\ 0 \\ 0 \\ 1\end{smallmatrix}\right]\!\!\left(2\mathbf{Z}\right)\ 
\theta\!\left[\begin{smallmatrix} 0\\ 1 \\ 1 \\ 0\end{smallmatrix}\right]\!\!\left(2\mathbf{Z}\right)\ 
\theta\!\left[\begin{smallmatrix} 1\\ 1 \\ 1 \\ 1\end{smallmatrix}\right]\!\!\left(2\mathbf{Z}\right)\ 
\right)^2\equiv\left[\Delta_{3/2}(\mathbf{Z})\right]^2\ .
\end{equation}
This is a known modular form with character of weight three at level four. For instance, see Aoki-Ibukiyama\cite{Aoki:2005}, where this is called $f_3$. Our procedure clearly provides a Borcherds product formula for it. 
Further, we will see in a later section that $\Delta_{3/2}(\mathbf{Z})$ as defined above appears as the denominator formula of a Borcherds Kac-Moody superalgebra, $\mathcal{G}_4$, in line with the notation introduced in our earlier paper\cite{Govindarajan:2008vi}.

\subsection{Product formula for $\widetilde{\Phi}_3(\mathbf{Z})$}

The product formula for  $\widetilde{\Phi}_3(\mathbf{Z})$ is 
\begin{align}
 \widetilde{\Phi}_3(\mathbf{Z}) = q^{1/4} r s  \prod_{a}^{3} \prod_{\substack{\ell,m\in \BZ,\\ n \in \BZ +
\tfrac{a}{4}}} \Big(1- q^nr^\ell s^m\Big)^{\sum_{b=0}^3 \omega^{-bm}
c^{(a,b)}(4nm -\ell^2)}
\end{align}
where $\omega=\exp(\tfrac{2\pi\imath}3)$ is a cube root of unity, 
$c^{(a,b)}(4nm- \ell^2)$ are the Fourier coefficients of the twisted elliptic
genera, $F^{(a,b)}(z_1,z_2)$.

As in the case of $\Phi_3(\mathbf{Z})$, $ \widetilde{\Phi}_3(\mathbf{Z}) $  can also be written as the square of the product of three even genus-two theta constants. By using the modular properties of the even genus-two theta constants, one obtains 
\begin{equation}
\widetilde{\Phi}_3(\mathbf{Z})=\left(\frac14\ \theta\left[\begin{smallmatrix} 0\\ 0 \\ 1 \\ 1\end{smallmatrix}\right]\!\!\left(\mathbf{Z}'\right)\ 
\theta\!\left[\begin{smallmatrix} 1\\ 1 \\ 0 \\ 0\end{smallmatrix}\right]\!\!\left(\mathbf{Z}'\right)\ 
\theta\!\left[\begin{smallmatrix} 1\\ 1 \\ 1 \\ 1\end{smallmatrix}\right]\!\!\left(\mathbf{Z}'\right)\ 
\right)^2\equiv\left[\widetilde{\Delta}_{3/2}(\mathbf{Z})\right]^2\ .
\end{equation}
where $\mathbf{Z}'= \left(\begin{matrix}\tfrac12  z_1 & z_2 \\ z_2 & 2 z_3 \end{matrix} \right)$. We have defined $\widetilde{\Delta}_{3/2}(\mathbf{Z})$ as the `square-root' of 
$\widetilde{\Phi}_3(\mathbf{Z})$ -- this will turn out to be given by the denominator formula of a Borcherds Kac-Moody superalgebra as we will discuss next.

\subsection{Integrality properties of the modular forms}\label{integrality}

One can prove that all the exponents that appear in the product formulae for $\Phi_3(\mathbf{Z})$ and $ \widetilde{\Phi}_3(\mathbf{Z})$ are all even integers.  One can show that the following expressions
$$
\left[4 A(\tau,z)-B(\tau,z)\right]/12\ , \quad [E_2(\tau)-1]/24\  \textrm{ and } [E_4(\tau)-1]/8{}
$$ 
all have integral Fourier coefficients\cite[see appendix A]{Cheng:2008kt}. A straightforward but tedious computation then shows that all exponents are even integers.

On the sum side, the integrality of coefficients in the Fourier expansion follows from the integrality properties of the genus-two theta constants.

\section{BKM Lie superalgebras}

Having constructed the Siegel modular forms $\Phi_3(\mathbf{Z})$ and $\widetilde{\Phi}_3(\mathbf{Z})$ we use them to explore the possibility of the existence of an algebraic structure to the $\tfrac14$-BPS states in the CHL model with a $\BZ_4$ orbifolding. For prime $N$ of the orbifolding group $\BZ_N$, the $\tfrac14$-BPS states have been found to have an underlying BKM  Lie superalgebra sturucture, so it is natural to ask if a similar structure exists for non-prime $N$. Cheng and Dabholkar have argued, based on general considerations, that the Siegel modular forms generating the dyon spectrum for $N>3$ will not have an underlying BKM Lie superalgebra structure\cite{Cheng:2008kt}. The modular form $\widetilde{\Phi}_3(\mathbf{Z})$, however, has not been constructed before. Hence a direct and explicit demonstration of the above argument has not been carried out. Having explicitly constructed the modular form in question, we proceed to show that there is indeed a BKM Lie superalgebra for both modular forms.

As for the case of prime $N$, there are two BKM Lie superalgebras associated to the `square roots' of the two genus-two Siegel modular forms $\widetilde{\Phi}_3(\mathbf{Z})$ and $\Phi_3(\mathbf{Z})$, denoted by $\widetilde{\Delta}_{3/2}(\mathbf{Z})$ and $\Delta_{3/2}(\mathbf{Z})$ respectively. To construct the BKM Lie superalgebras from the modular forms $\Delta_{3/2}(\mathbf{Z})$ and $\widetilde{\Delta}_{3/2}(\mathbf{Z})$ we adopt the procedure used in \cite{Nikulin:1995,Govindarajan:2008vi}. We compare our findings with the observations made by Cheng and Dabholkar\cite{Cheng:2008kt} with regards the roots of the BKM Lie superalgebra for $N>3$.

\subsection{Denominator formulae}
The Weyl-Kac-Borcherds (WKB) denominator formula is a special case of the more general WKB character formula for Lie algebras which gives the characters of integrable highest weight representations of BKM Lie superalgebras. The WKB character formula applied to the trivial representation gives the WKB denominator formula. Let $\mathfrak{g}$ be a BKM Lie superalgebra and $\mathcal{W}$ its Weyl group. Let $L_+$ denote the set of positive roots of the BKM Lie superalgebra and $\rho$ the Weyl vector. Then, the WKB denominator identity for the BKM Lie superalgebra $\mathfrak{g}$ is
\begin{equation}
\label{WKB}
\prod_{\alpha \in L_+} (1- e^{- \alpha})^{\textrm{mult}(\alpha)} = e^{-\rho}\
\sum_{w \in \mathcal{W}} (\det w)\  w(e^{\rho} \sum_{\alpha \in L_+}
\epsilon (\alpha) e^{\alpha} )\  , 
\end{equation}
where  mult$(\alpha)$ is the multiplicity of a root $\alpha \in L_+$. In the above equation, det($w$) is defined to be $\pm1$ depending on whether $w$ is the product of an even or odd number of reflections and $\epsilon (\alpha)$ is defined to be $(-1)^n$ if $\alpha$ is the sum of $n$ pairwise independent, orthogonal imainary simple roots, and $0$ otherwise.  In the case of BKM Lie superalgebras the roots appear with graded multiplicity -- fermionic roots appear with negative multiplicity while bosonic roots appear with positive multiplicity. The reader is referred to \cite{Borcherds:1990, Ray:2006} for a discussion on the denominator identity for BKM Lie superalgebras in general, and to \cite{Nikulin:1995,Govindarajan:2008vi} for a discussion in relation to the above problem, in particular.

In the sequel, our strategy  will be to use the method of \cite{Nikulin:1995} and \cite{Govindarajan:2008vi} to construct the BKM Lie superalgebras whose denominator identities are equal to  $\widetilde{\Delta}_{3/2}(\mathbf{Z})$ and $\Delta_{3/2}(\mathbf{Z})$. 
All the Fourier coefficients of $\widetilde{\Delta}_{3/2}(\mathbf{Z})$ and $\Delta_{3/2}(\mathbf{Z})$  are integral as discussed in Sec. \ref{integrality}.
Here we briefly recall the steps involved in it. Having obtained the product representations of $\widetilde{\Delta}_{3/2}(\mathbf{Z})$ and $\Delta_{3/2}(\mathbf{Z})$ we interpret these as the product side (L.H.S.) of the denominator identity \eqref{WKB}. Comparing with the above equation, this gives us the set of positive roots $\alpha$ of the BKM Lie superalgebra together with their multiplicities. All multiplicities in the product side are integral as the multiplicities in the product formulae are even integers as discussed in Sec. \ref{integrality}.
Also, expanding the modular form, we equate the expansion to the sum side (R.H.S.) of the denominator formula where each term is thought as coming from the Weyl reflection of a positive root with respect to an element of the Weyl group of the BKM Lie superalgebra. Thus, interpreting the modular form as the denominator formula, we can extract the positive roots and corresponding multiplicities, the set of simple roots, the Weyl group, the Weyl vector and from the above information, the Cartan matrix of the BKM Lie superalgebra. This procedure has been discussed in detail in the appendix D of \cite{Govindarajan:2008vi}  where the BKM Lie superalgebras that arise from  $\Delta_{k/2}(\mathbf{Z})$ for CHL orbifolds with prime $N$ have been derived. We now apply the above procedure to $\widetilde{\Delta}_{3/2}(\mathbf{Z})$ and $\Delta_{3/2}(\mathbf{Z})$ below.

\subsection{A BKM superalgebra for $\widetilde{\Delta}_{3/2}(\mathbf{Z})$}

Applying the above procedure to the expansion of $\widetilde{\Delta}_{3/2}(\mathbf{Z})$ we identify the factor $q^{1/8}r^{1/2}s^{1/2}$ with exp$(- \pi i (\rho^{(4)},\mathbf{Z}) )$. 
Let $(\delta_1,\delta_2,\delta_3)$ be three root vectors in hyperbolic space with norm given by the matrix\footnote{Recall that these are the simple real roots associated with the BKM Lie superalgebra whose denominator formula is $\Delta_5(\mathbf{Z})$\cite{Nikulin:1995}.}
\begin{equation}
\label{untwistedCartan}
A_{1,II}=\begin{pmatrix}
2 & -2 & -2 \\
-2 & 2 & -2 \\
-2 & -2 & 2
\end{pmatrix}\ .
\end{equation}
Using the identification (see appendix D1 in \cite{Govindarajan:2008vi})
 $$
e^{-\pi i (\delta_1,\mathbf{Z})}=q r\ ,\  e^{-\pi i (\delta_2,\mathbf{Z})}=r^{-1} \  
\textrm{ and }e^{-\pi i (\delta_3,\mathbf{Z})}=s r\ .
$$
we see that  the Weyl vector $\rho^{(4)}=\tfrac18\delta_1+\tfrac18\delta_2+\tfrac12\delta_3$. One can verify that $\rho$ is light-like, i.e., it has zero norm.  As discussed in \cite{Nikulin:1995,Cheng:2008fc,Govindarajan:2008vi}, $(\delta_1,\delta_2,\delta_3)$ can be written as $PGL(2,\mathbb{Z})$ matrices as follows:
\begin{equation}
\label{Cartan1}
  \delta_1 = \begin{pmatrix} 2 & 1 \\ 1 & 0 \end{pmatrix}\ , \ 
\delta_2= \begin{pmatrix} 0 & -1 \\ -1 & 0 \end{pmatrix}\ , \ 
\delta_3 = \begin{pmatrix} 0 & 1 \\ 1 & 2 \end{pmatrix}\ . \ 
\end{equation}
One also has $\rho^{(4)}=\begin{pmatrix} 1/4 & 1/2 \\ 1/2 & 1 \end{pmatrix}$ in agreement with the general formula given in ref. \cite[see Eq. 5.2]{Cheng:2008kt}.

Expanding  $\widetilde{\Delta}_{3/2}(\mathbf{Z})$ to about the first five thousand terms, we find the following terms (corresponding to real simple roots) appearing with multiplicity one -- there are infinitely more as we will prove later.
\begin{equation}
r^{-1}\ , \ qr \ , \ r s^4 \ , \ q r^7 s^{12}\ ,\ q^3 r^{17} s^{24} \ , \ q^3 r^7 s^4\ ,\ q^6 r^{17} s^{12}\ ,\ q^{10} r^{31} s^{24}\ . 
\end{equation}
These eight terms can be represented by the following eight $PGL(2,\mathbb{Z})$ matrices.
\begin{eqnarray}
\label{simpleroots}
\alpha_0 \equiv 
\begin{pmatrix}0& -1 \\ -1& 0 \end{pmatrix}, \quad \beta_0 \equiv \begin{pmatrix}2& 1 \\ 1& 0 \end{pmatrix}\ , \quad \beta_{-1} \equiv \begin{pmatrix}0& 1 \\ 1& 8 \end{pmatrix}\ , \nonumber \\ 
\alpha_1 \equiv \begin{pmatrix}2& 7 \\ 7& 24 \end{pmatrix}, \quad \beta_{-2} \equiv \begin{pmatrix}6& 17 \\ 17& 48 \end{pmatrix}\ , \quad \alpha_{-1} \equiv \begin{pmatrix}6& 7 \\ 7& 8 \end{pmatrix}\ , \\ \beta_1 \equiv \begin{pmatrix}12& 17 \\ 17& 24 \end{pmatrix}\ ,  \quad \alpha_{-2} \equiv \begin{pmatrix}20& 31 \\ 31& 48 \end{pmatrix}\ . \nonumber
\end{eqnarray}
Using the definition of the even genus-two theta constants, one can easily prove the following two identities.
\begin{enumerate}
\item Let $\mathbf{Z}'=\begin{pmatrix}z_1 & -z_2 \\ -z_2 & z_3\end{pmatrix}$. Then,
\begin{equation}\label{oddWeyl}
\widetilde{\Delta}_{3/2}(\mathbf{Z}')=-\widetilde{\Delta}_{3/2}(\mathbf{Z})\ .
\end{equation}
This implies that the modular form is an odd function under $r\rightarrow r^{-1}$ as can be seen easily in the Fourier expansion given in appendix D.
\item $\widetilde{\Delta}_{3/2}(\mathbf{Z})$ is invariant under the exchange $z_1 \leftrightarrow 4 z_3$. This implies that the modular form is an odd function under the exchange $q\leftrightarrow s^4 $ as can be seen easily in the Fourier expansion given in appendix D.
\end{enumerate}

We will now see if these results are compatible with expectations based on the walls of marginal stability for the $\mathbb{Z}_4$-orbifold.

\subsubsection{Walls of marginal stability}

Sen has analyzed the walls of marginal stability in the axion-dilaton plane (modelled by the upper-half plane with coordinate $\lambda$) by studying the decay of torsion one $\tfrac14$-BPS states into a pair of $\tfrac12$-BPS states\cite{Sen:2007vb}(see also\cite{Mukherjee:2007nc,Mukherjee:2007af}). We quote some of his results that are relevant for our considerations. Consider the following decay of a torsion one $\tfrac14$-BPS dyon into two $\tfrac12$-BPS dyons 
\begin{equation}
\begin{pmatrix} \mathbf{q}_e \\ \mathbf{q}_m\end{pmatrix}\longrightarrow
\begin{pmatrix}a d \ \mathbf{q}_e -b d\ \mathbf{q}_m \\ ca\ \mathbf{q}_e -cb\ \mathbf{q}_m \\ \end{pmatrix}
\oplus \begin{pmatrix} -bc\ \mathbf{q}_e +bd\ \mathbf{q}_m \\ -ac\ \mathbf{q}_e + ad\ \mathbf{q}_m \\ \end{pmatrix}\ ,
\end{equation}
where the kinematics of the decay imply that the integers $a,b,c,d$ are such that\cite{Sen:2007vb}
\begin{enumerate}
\item $ad-bc=1$.
\item The equivalence relation $(a,b,c,d)\sim (a \sigma^{-1},b \sigma^{-1}, c \sigma, d\sigma)$ with $\sigma\neq0$.
\item Exchanging the two decay products implies the equivalence under:\\[4pt]
\centerline{ $(a,b,c,d)\rightarrow (c,d,-a,-b)$.}
\item Charge quantization requires $ad, bd, bc\in \mathbb{Z}$ and $ac\in N\mathbb{Z}$.
\end{enumerate}
One can show that by suitable use of the equivalences given above, one can always choose $\begin{pmatrix} a & b \\ c & d \end{pmatrix}\in \Gamma_1(N)$ for $N=2,3,4$.

This decay occurs across real codimension one walls in the upper-half plane -- the $\tfrac14$-BPS state decays into two $\tfrac12$-BPS states as one moves across the wall. 
In the upper-half plane, these  walls are circular arcs determined by the equation\cite{Sen:2007vb,Mukherjee:2007nc}
\begin{equation}
\left[\textrm{Re}(\lambda)-\tfrac{ad+bc}{2ac}\right]^2 + \left[\textrm{Im}(\lambda)+ \tfrac{\mathcal{E}}{2ac}\right]^2 = \tfrac{1+\mathcal{E}^2}{4a^2c^2}\ ,
\end{equation}
where $\mathcal{E}$ is a real function of all other moduli $M$. It is easy to see the arcs intersect the real $\lambda$ axis at the points $\tfrac{b}{a}$ and $\tfrac{d}{c}$ for any $\mathcal{E}$. When $\mathcal{E}=0$, the arcs are semi-circles centred on the real $\lambda$-axis with radius $\tfrac1{2ac}$. When $\mathcal{E}\neq0$, the center of the circle moves into the interior of the upper half plane with radius also increasing -- all this with the intercepts on the real axis remaining unchanged.When either $a=0$ or $c=0$, the circles become straight lines perpendicular to the real axis for $\mathcal{E}=0$ and making a suitable angle for $\mathcal{E}\neq0$. For simplicity, we restrict the discussion in the sequel to the case when $\mathcal{E}=0$ -- as the sole effect on non-zero $\mathcal{E}$ is to `deform' the semi-circles into circular arcs.

\begin{figure}[h]
\centering
\includegraphics[height=2in]{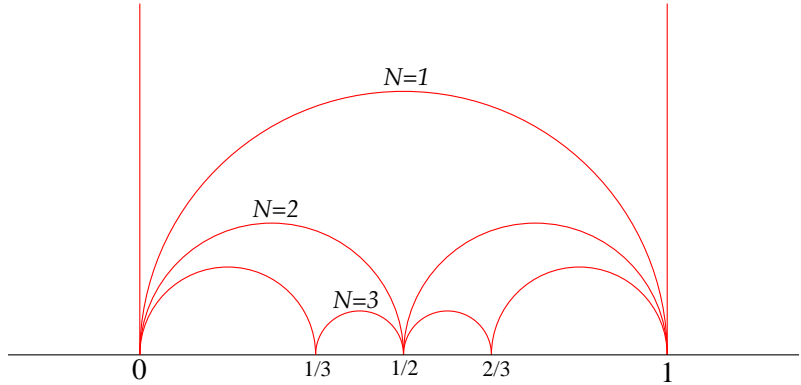}
\caption{Fundamental domains/Weyl chamber for $N=1,2,3$}
\label{weylchamber123}
\end{figure}

A fundamental domain is constructed by first restricting the value of Re$(\lambda)$ to the interval $[0,1]$. The straight lines Re$(\lambda)=0,1$ correspond to two walls of marginal stability. Next, one looks for the largest semi-circle with one end at $\lambda=0$ on the real axis that is compatible with the quantization of charges. This semi-circle intersects the real axis at some point in the interval $[0,1]$ -- this turns out to be at $1/N$. The procedure is then (recursively) repeated by looking for another semi-circle with one end at $1/N$ till one hits the mid-way point $1/2$. A similar procedure is done starting with the largest semi-circle with one end on the point $\lambda=1$ on the real axis. One obtains the following set of points for $N=1,2,3$:
\begin{equation}
(\tfrac01,\tfrac11)\ ,\quad (\tfrac01,\tfrac12,\tfrac11)\ , \quad (\tfrac01,\tfrac13, \tfrac12,\tfrac23,\tfrac11)\ .
\end{equation}
A fundamental domain is then given by restricting to the region bounded by these semi-circles and the two walls connecting $\lambda=0,1$ to infinity. The two straight lines may be included by adding the `points' $\tfrac{-1}0$ and $\tfrac10$. 
The fundamental domains are given in Figure \ref{weylchamber123}.

\begin{figure}[h]
\centering
\includegraphics[height=2.5in]{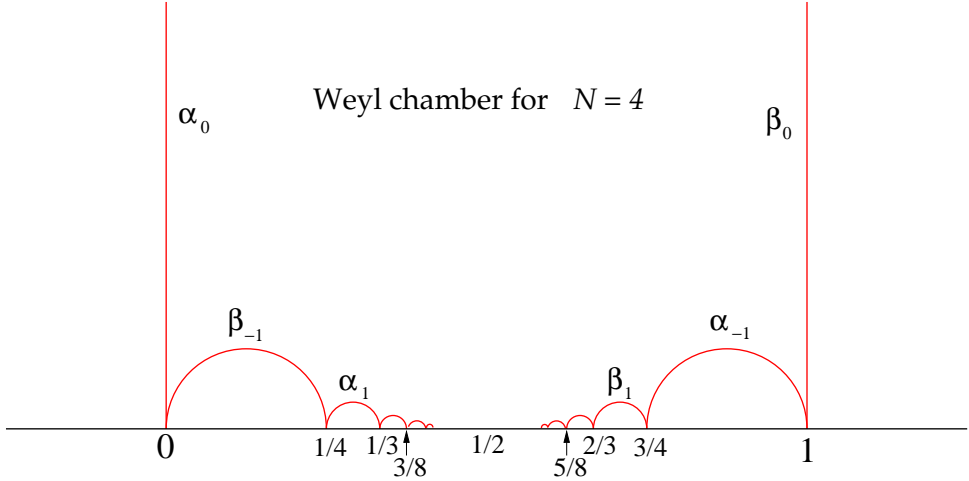}
\caption{The fundamental domain/Weyl chamber for $N=4$ is bounded by an infinite number of semi-circles as the BKM Lie superalgebra has infinite  number of real simple roots. Each of the  semi-circles indicated represent real simple roots that appear with multiplicity one in the sum side of the denominator formula. Note that the diameter of the semi-circles are reducing as one gets closer to $\tfrac12$. The point $\tfrac12$ is approached as a limit point of the infinite sequence of semi-circles.}
\label{weylchamber4}
\end{figure}

For $N>3$, this picture does not terminate -- one needs an infinite number of semi-circles to obtain a closed domain. For $N=4$, we find the following sequence on (using Sen's method)
\begin{equation}
(\tfrac01,\tfrac14,\tfrac13,\tfrac38,\tfrac25,\ldots,\tfrac{-2n+1}{-4n},\tfrac{-n}{-2n-1},\ldots,\tfrac12,\ldots,\tfrac{n+1}{2n+1},\tfrac{2n+1}{4n}\ldots,\tfrac35,\tfrac58,\tfrac23,\tfrac34,\tfrac11)\ .
\end{equation}
Let $\alpha_n$ denote  the semi-circle with intercepts $\big(\tfrac{2n-1}{4n},\tfrac{n}{2n+1}\big)$
 and $\beta_n$ the semi-circle with intercepts $\big(\tfrac{n+1}{2n+1},\tfrac{2n+1}{4n}\big)$  for all $n\in\mathbb{Z}$. Note that $\alpha_0$ and $\beta_0$ represent the two straight lines at Re$(\lambda)=0,1$ respectively.
The fundamental domain corresponding to the above sequence is depicted in Figure \ref{weylchamber4}. It may be thought of as a regular polygon with infinite edges with the infinite dimensional dihedral group, $D_\infty=\mathbb{Z}\rtimes \mathbb{Z}_2$, as its symmetry group. $D^{(1)}_\infty$ is generated by two generators: a reflection $y$ 
and a shift $\gamma$ given by:
\begin{equation}
\label{Dinfinity}
y:\quad  \alpha_n \rightarrow \alpha_{-n}\ ,\ \beta_n \rightarrow \beta_{-n-1}\quad \textrm{and}\quad  \gamma: \alpha_n \rightarrow \alpha_{n+1}\ ,\ 
\beta_n \rightarrow \beta_{n-1}\ ,
\end{equation}
satisfying the relations $y^2=1$ and $y\cdot \gamma\cdot y=\gamma^{-1}$. There is a second $\BZ_2$ generated by $\delta$ defined as follows:
\begin{equation}
\delta:\quad \alpha_n \longleftrightarrow \beta_n\ .
\end{equation}
The transformations $(\gamma,\delta)$ generate another dihedral group that we denote by $D^{(2)}_\infty$.

\subsubsection{Walls of a Weyl chamber}

We have just seen that the fundamental domain for the $\mathbb{Z}_4$-orbifold  in the $\lambda$-plane was bounded by an infinite number of edges. (Cheng and Verlinde\cite{Cheng:2008fc} and Cheng and Dabholkar\cite{Cheng:2008kt} have shown the for $N=1,2,3$, this fundamental domain is the Weyl chamber of a family of rank-three BKM Lie superalgebras.) Each wall (edge) of the fundamental domain is identified with a real simple root of the BKM Lie superalgebra. Recall that each wall corresponds to a pair of rational numbers $(\tfrac{b}a,\tfrac{d}c)$. This is related to a real simple root $\alpha$ of the BKM Lie superalgebra as follows:
\begin{equation}
(\tfrac{b}a,\tfrac{d}c) \leftrightarrow \begin{pmatrix}a & b \\ c & d \end{pmatrix} \leftrightarrow
\alpha=\begin{pmatrix} 2bd & ad + bc \\ ad + bc & 2 ac \end{pmatrix} \ ,
\end{equation}
with $ac\in N\mathbb{Z}$ and $ad,bc,bd\in \mathbb{Z}$. The norm of the root is\cite{Cheng:2008fc} 
$$
-2\det(\alpha)=2(ad-bc)^2=2\ .
$$ 
The Cartan matrix, $A^{(N)}$, is generated by the matrix of inner products among all real simple roots. For instance, $A^{(1)}=A_{1,II}$ defined in Eq. \eqref{Cartan1}.

The `square root' of the modular form $\widetilde{\Phi}_k(\mathbf{Z})$ that generates dyon degeneracies, $\widetilde{\Delta}_{k/2}(\mathbf{Z})$, is related to the Weyl-Kac-Borcherds denominator formula via its additive and multiplicative lifts. Finally, the extended S-duality group is given by\footnote{The extended S-duality group is defined by including a $\mathbb{Z}_2$ parity operation to the S-duality group $\Gamma_1(N)$. For $N=1$, this is the group $PGL(2,\mathbb{Z})$\cite{Cheng:2008fc}.}
\begin{equation}
\mathcal{W}(A^{(N)})\rtimes D_N\ ,
\end{equation}
where $\mathcal{W}(A^{(N)})$ is the group  generated by Weyl reflections of all the real simple roots\footnote{This is equivalent to the Coxeter group generated by the Cartan matrix $A^{(N)}$.} and $D_N$ is the dihedral group that is the symmetry group of the polygon corresponding to the Weyl chamber.

As we will now show, the correspondence goes through for $N=4$ even though the number of real simple roots is infinite. Ordering the real simple roots into an infinite-dimensional vector 
$$
\mathbf{X}=(\ldots, x_{-2},x_{-1},x_{0},x_{1},x_2,x_3,\ldots)=( \ldots, \alpha_1, \beta_{-1}, \alpha_0, \beta_0,\alpha_{-1},\beta_1,\ldots)\ .$$
Equivalently, let
\begin{equation}
x_m = \left\{\begin{array}{ll} \alpha_{-m/2}\ , & m\in 2\mathbb{Z}\\
                                            \beta_{(m-1)/2}\ , & m \in 2\mathbb{Z}+1\ .
                                            \end{array} \right.
\end{equation}
The Cartan matrix is given by the matrix of inner products $a_{mn}\equiv \langle x_n , x_m \rangle$ and is given by the infinite-dimensional matrix:
\begin{equation}\label{infiniteCartan}
A^{(4)} = (a_{nm})\quad \textrm{where}\quad 
a_{nm}= 2 -4(n-m)^2\  ,
\end{equation}
with $m,n\in \mathbb{Z}$. It is easy to show that the following family of vectors are eigenvectors of the Cartan matrix with zero eigenvalue.
\begin{equation}
\begin{pmatrix} \vdots \\ ~1 \\ -3 \\ ~3 \\ -1 \\ \vdots \end{pmatrix}\ ,
\end{equation}
with the vertical dots ($\vdots$) indicating a semi-infinite sequence of zeros. One can show that $A^{(4)}$ has rank three. The Weyl vector $\rho^{(4)}$ satisfies
\begin{equation}
\langle \rho^{(4)}, x_m\rangle =-1\ ,\ \forall m \ .
\end{equation}

\subsubsection*{$D^{(2)}_\infty$-invariance of $\widetilde{\Delta}_{3/2}(\mathbf{Z})$}

It remains to be proven that $\widetilde{\Delta}_{3/2}(\mathbf{Z})$ gives rise to the denominator identity for this BKM Lie superalgebra. The analysis of the sum side of the expansion has given eight real simple roots (to the order we checked) that we listed in Eq. \eqref{simpleroots} which all belong to the set $\mathbf{X}$ of real simple roots.We will now show that it contains all the real simple roots that one expects from the study of the walls of marginal stability. The $D^{(2)}_\infty$-generators  $\gamma$ and $\delta$ act on the roots $x_m$ written as a $2\times2$ matrix as follows:
\begin{align}
\gamma:\  &x_m \longrightarrow \begin{pmatrix} 1 & -1 \\ 4 & -3 \end{pmatrix}\cdot x_m \cdot \begin{pmatrix} 1 & -1 \\ 4 & -3 \end{pmatrix}^{\textrm{T}}\ ,\\
\delta:\  &x_m \longrightarrow \begin{pmatrix} -1 & 1 \\ 0 & 1 \end{pmatrix}\cdot x_m \cdot \begin{pmatrix} -1 & 1 \\ 0 & 1 \end{pmatrix}^{\textrm{T}}\ .
\end{align}
The matrix $\gamma$ is denoted  by $\gamma^{(4)}$ in \cite{Cheng:2008kt}. In appendix \ref{lastone}, we showed the invariance of the square of $\widetilde{\Delta}_{3/2}(\mathbf{Z})$ under the  symmetry generated by the embedding of $\gamma$ and $\delta$ into $G_0(4)\in Sp(2,\BZ)$. This implies that under the action of $\gamma$ and $\delta$,  
$$
\widetilde{\Delta}_{3/2}(\mathbf{Z})\rightarrow \pm\  \widetilde{\Delta}_{3/2}(\mathbf{Z})\ .
$$
One can show that the sign must be $+1$ by observing that any pair of terms in the Fourier expansion of $\widetilde{\Delta}_{3/2}(\mathbf{Z})$ related by the action of $\gamma$ ($\delta$ resp.) appear with the same Fourier coefficient. For instance, the terms associated with the two simple roots $\alpha_0$ and $\beta_0$ related by the action of $\delta$ appear with coefficient $+1$. Similarly, the terms associated with the real simple roots $\beta_0$ and $\beta_{-1}$ related by a $\gamma$-translation also appear with coefficient $+1$. Thus, we see that $\widetilde{\Delta}_{3/2}(\mathbf{Z})$ is invariant under the full dihedral group $D^{(2)}_\infty$. This provides an \textit{all-orders} proof that the infinite number of real simple roots given by the vector $\mathbf{X}$ all appear in the Fourier expansion of $\widetilde{\Delta}_{3/2}(\mathbf{Z})$.

The $q\rightarrow s^4$ symmetry of the modular form is equivalent to the symmetry generated by the dihedral generator, $y$, as defined in Eq. \eqref{Dinfinity}.

\subsubsection*{Weyl transformation of $\widetilde{\Delta}_{3/2}(\mathbf{Z})$}

The transformation $r\rightarrow r^{-1}$ is the Weyl reflection about the root $\alpha_0$ and as discussed earlier (see Eq. \eqref{oddWeyl}), the modular form is odd under the Weyl reflection. One has
\begin{equation}
w_{\alpha_0}\cdot \mathbf{Z} = \begin{pmatrix} 1 & 0 \\ 0 & -1 \end{pmatrix}^{\textrm{T}} \cdot \mathbf{Z} \cdot \begin{pmatrix} 1 & 0 \\ 0 & -1 \end{pmatrix}\ .
\end{equation}
The reflection due to any other elementary Weyl reflection will also have the same sign. We repeat an argument from the appendix A of \cite{Cheng:2008kt} to show this. First, the reflection due to $\alpha_0$ is represented by the matrix $w_0\equiv \begin{pmatrix} 1 & 0 \\ 0 & -1 \end{pmatrix}$. The action on $\mathbf{Z}$ is equivalent to $Sp(2,\BZ)$ action by the matrix\cite{Nikulin:1995}
$$
M=\begin{pmatrix} (w_0^{-1})^{\textrm{T}}& 0 \\ 0 &w_0  \end{pmatrix}\ ,
$$
The minus sign due to the Weyl reflection implies that the character, $v(M)$,  associated with the modular form $\widetilde{\Delta}_{3/2}(\mathbf{Z})$ is such that $v(M)=-1$.
Next, any other elementary Weyl reflection, $w$,  must be conjugate to $w_0$ -- this is a consequence of the dihedral symmetry, $D^{(2)}_\infty$. Hence, one has $w= s\cdot w_0\cdot  s^{-1}$ for some invertible matrix $s$. It follows that the character associated with the Weyl reflection $w$ is the same as that for $w_0$. In others, $\widetilde{\Delta}_{3/2}(\mathbf{Z})$ is odd under \textit{all} elementary reflections. Hence one has
\begin{equation}
\widetilde{\Delta}_{3/2}(w\cdot \mathbf{Z})=\det(w)\  \widetilde{\Delta}_{3/2}(\mathbf{Z})\ .
\end{equation}

We thus see that the extended S-duality group for $N=4$ is given by\footnote{The generator $y$ is not realized as an element of a level $4$ subgroup of $PGL(2,\BZ)$ and thus is not an element of the extended S-duality group. This is similar to what happens for $N=2,3$\cite{Cheng:2008kt}. We thank M. Cheng and A. Dabholkar for  useful email correspondence.}
\begin{equation}
\mathcal{W}(A^{(4)})\rtimes D^{(2)}_\infty\ ,
\end{equation}
where $\mathcal{W}(A^{(4)})$ is the Coxeter group generated by the reflections by all real simple roots $x_m$ and $D^{(2)}_\infty$ is the infinite dimensional dihedral group generated by $\gamma$ and $\delta$.

\subsection{A BKM superalgebra for $\Delta_{3/2}(\mathbf{Z})$}

Starting from the product expansion for $\Delta_{3/2}(\mathbf{Z})$ we now apply the same procedure to find the BKM Lie superalgebra associated to it. From the expansion, we identify the common factor $q^{1/2}r^{1/2}s^{1/2}$ with exp$(- \pi i (\rho,z) )$ and thus, as before for the case of prime $N$, the Weyl vector $\rho$ does not change upon orbifolding for the BKM Lie superalgebras coming from the $\Delta_{k/2}(\mathbf{Z})$. We also observe that the three real simple roots remain unchanged as before\cite{Govindarajan:2008vi}. The imaginary roots remain unchanged as well, but their multiplicities are changed by the orbifolding. For $\Delta_{k/2}(\mathbf{Z}) = (\Phi_k(\mathbf{Z}))^{1/2}$ for prime $N$, we recall that the BKM Lie superalgebras $\mathcal{G}_N$ were all given by the same Cartan matrix, $A_{1,II}$ (given in Eq. \eqref{untwistedCartan}) and had the same set of real simple roots, Weyl group, Weyl vector, and imaginary roots. The orbifolding only changed the multiplicities of the imaginary roots for different values of $N$. We see that the same pattern continues to hold for the BKM Lie superalgebra even when $N$ is non-prime for $\Delta_{3/2}(\mathbf{Z})$.

\subsection{Physical interpretation of the superalgebras $\Delta_{k/2}(\mathbf{Z})$}

We have seen that the BKM Lie superalgebas associated with the modular forms $\Delta_{k/2}(\mathbf{Z})$ that appear for $N=1,2,\ldots, 5$ all have the same walls of marginal stability but differ in the multiplicities of the imaginary simple roots. Since the appearance of an earlier version of this paper, Sen has shown that the Siegel modular forms associated with a twisted helicity trace index are indeed the square of the modular forms $\Delta_{k/2}(\mathbf{Z})$ -- these compute the degeneracies of a sub-sector of dyons in heterotic string theory on $T^6$\cite{Sen:2009md}. The S-duality group and the walls of marginal stability for dyons in this sub-sector remain unchanged. Thus, it appears that the BKM Lie superalgebras $\mathcal{G}_N$ play the role of $\mathcal{G}_1$ that appears for the untwisted trace\footnote{We thank Atish Dabholkar for drawing our attention to the possibility of this interpretation.}.  It is easy to see that the counting of  $\tfrac12$-BPS dyons in this sub-sector are also captured by the same 
multiplicative $\eta$-products that appear in the $\mathbb{Z}_N$ orbifold.

\subsection{Observations}

We have seen that the BKM Lie superalgebra for $\Delta_{3/2}(\mathbf{Z})$ exists and exhibits the pattern in properties observed for the series $\mathcal{G}_N$ and the BKM Lie superalgebra interpretation for the modular form $\widetilde{\Delta}_{3/2}(\mathbf{Z})$ does appear to fit the denominator formula for a BKM Lie superalgebra. This algebra does \textit{not} make the list of rank-three Lorentzian Kac-Moody algebras of Gritsenko and Nikulin\cite{Gritsenko:2002}(see also \cite{GritsenkoNikulinI,GritsenkoNikulinII}) as it violates a finiteness condition (on the volume of the Weyl chamber) imposed by them. A closely related issue is that $\widetilde{\Delta}_{3/2}(\mathbf{Z})$ is a meromorphic modular form\cite{Jatkar:2005bh}.  Nevertheless, we claim that there is indeed a BKM Lie superalgebra with infinitely many real simple roots  (i.e., it is of parabolic type). Indeed, such an example has been already considered in \cite{GritsenkoNikulinII} where a BKM Lie superalgebra has been associated with a single genus-two theta constant. As we discuss in the sequel, this example is associated with the one of the two distinct product orbifolds $\mathbb{Z}_4\times \mathbb{Z}_4$. 

\section{Generalizations}

\subsection{Product Groups}

The multiplicative $\eta$-products listed in Table \ref{Dummit} provide us with the generating function of $\tfrac12$-BPS states for all  type II orbifolds of $K3\times T^2$ that involve product groups $\mathbb{Z}_n\times \mathbb{Z}_m$ acting as Nikulin involutions on $K3$ and shifts of order $n$ and $m$ on the two circles that form the $T^2$. For instance, the $\mathbb{Z}_4\times \mathbb{Z}_4$-orbifold has cycle shape $4^6$ and hence the corresponding $\eta$-product is $\eta(4\tau)^6$. 

However, we need to obtain the seed for the additive lift. As the shift now acts on both the circles, it is not  straightforward to extend the microscopic computation of David-Sen using the 4d-5d lift\cite{David:2006yn,Gaiotto:2005gf}. Recall that in this approach, the spatial $R^3$ and  $\widetilde{S}^1$   was replaced by  the Taub-NUT geometry with NUT charge $1$. The natural replacement for the situation where the $\widetilde{S}^1$ has a shift of order $m$ (in the type IIB picture) is to choose Taub-NUT geometry with NUT charge $m$\cite{Gaiotto:2005gf,Dabholkar:2008zy}. Then, near $r=\infty$ the geometry is that of $\mathbb{R}^3\times \widetilde{S}^1/\mathbb{Z}_m$ where the orbifold action acts as a shift of order $m$. This is the $4d$ geometry. Near $r=0$, the geometry is of the form $\mathbb{R}^4/\mathbb{Z}_m$, this is the $5d$-geometry. When $n\neq m$, it appears that there are two different D1-D5 configurations where the roles of the $\mathbb{Z}_n$ and $\mathbb{Z}_m$ groups are interchanged.

In this setup, the other contribution to the additive seed arises from the center of mass motion of the D1-D5 branes in Taub-NUT space with NUT charge $m$. In the small $r$ limit, the Taub-NUT geometry reduces to $\mathbb{C}^2/\mathbb{Z}_m$. Combining it with the corresponding $\mathbb{Z}_m$ action on the $K3$, we see that the singularity is locally $\mathbb{R}^8/\mathbb{Z}_m$. For the values of $m$ that occur, the singularity is terminal\cite{Morrision:1984,Font:2004et} and there are \textit{no} massless modes in the twisted sectors\cite{Dasgupta:1995zm,Sen:1996zq,Roy:1996ca,Font:2004et}. Thus, the index gets contributions only from the untwisted sector and hence the center of mass contribution is \textit{independent} of the NUT charge. Thus in the type IIB frame, the additive seed is given by 
\begin{equation}
\frac{\vartheta_1(n \tau,z)^2}{\eta(n \tau)^6}\times g_\rho(\tau)\ .
\end{equation}
Thus, the additive seed for the $\mathbb{Z}_2\times \mathbb{Z}_2$ orbifold is given the following Jacobi form
\begin{equation}
\phi_{4,1}(\tau,z)=\vartheta_1(2 \tau,z)^2 \times \eta(2\tau)^{6}\ .
\end{equation}

We need to follow the chain of dualities to go to the heterotic frame. However, in the heterotic frame, the shift (of order $m$) is not  a momentum shift but a winding shift. Thus, there are two distinct orbifolds, one corresponding to a momentum shift in the type IIB frame (the one we just considered) and the one corresponding to a momentum shift in the heterotic frame. The corresponding threshold computation has been carried out by Banerjee, Jatkar and Sen\cite{Banerjee} for the $\mathbb{Z}_2\times \mathbb{Z}_2$-orbifold and the modular form is given by\footnote{We thank Dileep Jatkar for discussions regarding this work and help in sorting out some confusing aspects.}
\begin{equation}
\Phi_4(\mathbf{Z})= \frac{\Delta_{3}(\mathbf{Z})^3}{\Delta_5(\mathbf{Z})}\ ,
\end{equation}
where $\Delta_5(\mathbf{Z})$ and $\Delta_3(\mathbf{Z})$ are the square roots of the modular forms appearing in the $N=1,2$ models. This modular form can be obtained as the additive lift of the Jacobi form
\begin{equation}
\widehat{\phi}_{4,1}(\tau,z)=\frac{\vartheta_1( \tau,z)^2}{\eta( \tau)^6}\times \eta(2\tau)^{12}\ .
\end{equation}
We conjecture that the Jacobi form for additive lift for the $\mathbb{Z}_n\times \mathbb{Z}_m$-orbifolds in the heterotic frame is given by
\begin{equation}
\widehat{\phi}_{k,1}(\tau,z)=\frac{\vartheta_1( \tau,z)^2}{\eta( \tau)^6}\times g_\rho(\tau)\ ,\end{equation}
where the weight $k$ and cycle shape $\rho$ are as given in Table \ref{Dummit}.

\subsubsection{A classical formula}

The additive seed for the $\mathbb{Z}_4\times \mathbb{Z}_4$-orbifold in the type IIB frame is given by
\begin{equation}
\phi_{1,1}(\tau,z)=\vartheta_1(4\tau,z)^2\ .
\end{equation}
Note that the $\eta$-functions have cancelled out! Let us denote by $\Phi_1(\mathbf{Z})$ the genus-two modular form at level $4$ given by the additive lift. One can show that it can be written as the square of a single even genus-two theta constant.
\begin{equation}
\Phi_1(\mathbf{Z})= \left(\frac12 \theta\!\left[\begin{smallmatrix} 1\\ 1 \\ 1 \\ 1\end{smallmatrix}\right]\!\!\left(\mathbf{Z}'\right)\ 
\right)^2\equiv\left[\Delta_{1/2}(\mathbf{Z})\right]^2\ ,
\end{equation}
where $\mathbf{Z}'=\begin{pmatrix} 4 z_1 & 2 z_2 \\ 2z_2 & 4 z_3\end{pmatrix}$. This lift has been studied by Gritsenko and Nikulin who also have provided a Borcherds type product formula\cite{GritsenkoNikulinII}.  Consider the weak Jacobi form
\begin{equation}
2\frac{\vartheta_1(\tau,3z)}{\vartheta_1(\tau,z)}=\sum_{n,\ell} b(n,\ell)\ q^n r^\ell\ .
\end{equation}
Then,
\begin{equation}
\Phi_1(\mathbf{Z})=q r s \prod_{(n,\ell,m)>0} (1-q^{4n} r^\ell s^{4m})^{b(nm,\ell)}\ .
\end{equation}

The square root of this modular form is given by a single even genus-two theta constant\footnote{Gritsenko and Nikulin call this theta constant the most `odd' of the ten even genus-two theta constants\cite{GritsenkoNikulinII}.}. This is shown to be the denominator formula for a BKM superalgebra of parabolic type.  This superalgebra appears to be the analog of the affine KM algebra whose denominator formula is given by a single genus-one theta constant. In fact, we conjecture that the three genus-two modular forms labelled $\Delta_k$ ($k=2,1,1/2$) considered in \cite{GritsenkoNikulinII} are the `square-roots'  of the generating function of dyonic states in the $\mathbb{Z}_n\times \mathbb{Z}_n$ models (in the type IIB frame) with $N=2,3,4$ respectively. Further, it appears that BKM superalgebras associated with these models have the same Cartan matrix, $A^{(n)}$,  as the corresponding $\mathbb{Z}_n$ model.  We will discuss this  further in a future publication\cite{forthcoming2}.

\subsection{$\eta$-quotients and type II models}

There exist other $\mathcal{N}=4$ supersymmetric four-dimensional theories that can be obtained as $\mathbb{Z}_N$-orbifolds of the type II string compactified on $T^6$.  David, Jatkar and Sen have constructed genus-two modular forms that play a role similar to the ones considered for CHL orbifolds\cite{David:2006ru}. However, there are a few differences. Electrically charged $\tfrac12$-BPS states in this theory are counted by considering states of the superstring instead of the heterotic string. Unlike the case of CHL orbifolds,  where the index truly counted all states as there were no `fermionic' excitations for the bosonic sector of the heterotic string, the index only counts the difference between the `fermionic' and `bosonic' excitations. Further, there is no string tree level $R^2$ correction in type II models.\footnote{It is interesting to note that   the modular form that counts $\tfrac18$-BPS dyons in the type II string compactified on $T^6$ is given by the additive lift of the Jacobi form given above with $g_\rho(\tau)=1$\cite{Sen:2008sp}.}

The modular forms constructed by David, Jatkar and Sen in ref. \cite{David:2006ru} were for the $\mathbb{Z}_2$ and $\mathbb{Z}_3$-orbifolds and were generated by the additive lift of the Jacobi form
\begin{equation}
\frac{\vartheta_1( \tau,z)^2}{\eta( \tau)^6}\times g_\rho(\tau)\ ,
\end{equation}
where $g_\rho(\tau)$ is the quotient of products of $\eta$-functions which we call $\eta$-quotients. Like the CHL orbifolds, these quotients are associated with \textit{frame shapes}\cite{Mason:1985,MR1376550}. A  frame shape is a generalization of cycle shape where negative exponents are permitted. For instance, the  frame shape $\rho=1^{16} 2^{-8} $ is associated with the $N=2$ orbifold. Similarly, the $N=3$ orbifold is associated with the frame shape $\rho=1^9 3^{-3}$. 

It can be shown  that the genus-two modular forms that appear in the type II examples  can be written in terms of the CHL genus-two modular forms. Further, the work of Martin\cite{MR1376550} on multiplicative $\eta$-quotients enables us to extend the results for $N=2,3$ to include the other possibilities. This will be discussed in a forthcoming paper\cite{forthcoming}.

%There also appears to be a connection that relates such frame shapes with the Conway group $Co_0$ (also known as $.O$) -- the appearance of the Conway group is not clear to us at the moment. 

\section{Conclusion}

In this paper, we have completed the construction of the genus-two modular forms that count dyons in all $\mathbb{Z}_N$-orbifolds as well as given candidates for orbifolds involving product groups. These modular forms satisfy all the required consistency conditions. Given the additive and product formulae for the modular forms, we then proceeded to study the associated BKM Lie superalgebras whose Weyl-Kac-Borcherds denominator identity gives rise to the square-root of the modular forms. In particular, for the $\mathbb{Z}_4$ CHL orbifold, we have seen that there are two inequivalent BKM Lie superalgebras associated with the two modular forms $\Delta_{3/2}(\mathbf{Z})$ and $\widetilde{\Delta}_{3/2}(\mathbf{Z})$. Both these algebras satisfy the expectations from the considerations in earlier work\cite{Govindarajan:2008vi,Cheng:2008kt}.

\bigskip \bigskip

\noindent \textbf{Acknowledgments:} We would like to thank H. Aoki, A. Dabholkar, S. Gun,  D. Jatkar, Y. Martin and P.K. Tripathy  for useful discussions. We are grateful to Purusottam Rath for directing us to the paper by Dummit et. al.  KGK would like to thank S. Kalyana Rama and T.R. Govindarajan for their constant support. SG would like to thank the organizers of the Kanha String Meeting held in February 2009 for the opportunity to present some of these results.

\appendix

\section{Theta functions}
\subsection{Genus-one theta functions}\label{genusone}

The genus-one theta functions are defined by
\begin{equation}
\theta\left[\genfrac{}{}{0pt}{}{a}{b}\right] \left(z_1,z_2\right)
=\sum_{l \in \BZ} 
q^{\frac12 (l+\frac{a}2)^2}\ 
r^{(l+\frac{a}2)}\ e^{i\pi lb}\ ,
\end{equation}
where $a.b\in (0,1)\mod 2$ and $q=\exp(2\pi i z_1)$ and $r=\exp(2\pi i z_2)$.
 One has $\vartheta_1 
\left(z_1,z_2\right)\equiv\theta\left[\genfrac{}{}{0pt}{}{1}{1}\right] 
\left(z_1,z_2\right)$, $\vartheta_2 
\left(z_1,z_2\right)\equiv\theta\left[\genfrac{}{}{0pt}{}{1}{0}\right] 
\left(z_1,z_2\right)$, $\vartheta_3 
\left(z_1,z_2\right)\equiv\theta\left[\genfrac{}{}{0pt}{}{0}{0}\right] 
\left(z_1,z_2\right)$ and $\vartheta_4 
\left(z_1,z_2\right)\equiv\theta\left[\genfrac{}{}{0pt}{}{0}{1}\right] 
\left(z_1,z_2\right)$.

The transformations of $\vartheta_1(\tau,z)$ under modular transformations
is given by
\begin{eqnarray}
T:\qquad \quad\! \vartheta_1(\tau+1,z) &=& e^{i\pi/4}\
\vartheta_1(\tau,z)\ ,
\nonumber \\
S:\quad  \vartheta_1(-1/\tau,-z/\tau) &=& -\frac{1}{q^{1/2}r}\ e^{\pi
iz^2/\tau}\ \vartheta_1(\tau,z)\ .
\end{eqnarray}
with $q=\exp(2\pi i \tau)$ and $r=\exp(2\pi i z)$.

\subsection{Genus-two theta constants}\label{thetaconstants}

We define the genus-two theta constants as follows\cite{Nikulin:1995}:
\begin{equation}
\theta\left[\genfrac{}{}{0pt}{}{\mathbf{a}}{\mathbf{b}}\right]
\left(\mathbf{Z}\right)
=\sum_{(l_1, l_2)\in \BZ^2} 
q^{\frac12 (l_1+\frac{a_1}2)^2}\ 
r^{(l_1+\frac{a_1}2)(l_2+\frac{a_2}2)}\ 
s^{\frac12 (l_2+\frac{a_2}2)^2}\ 
e^{i\pi(l_1b_1+l_2b_2)}\ ,
\end{equation}
where $\mathbf{a}=\begin{pmatrix}a_1\\ a_2
\end{pmatrix}$,
$\mathbf{b}=\begin{pmatrix}b_1\\ b_2
\end{pmatrix}$,
and $\mathbf{Z}=\begin{pmatrix}z_1 & z_2 \\ z_2 &
z_3\end{pmatrix}\in \mathbb{H}_2$. 
Further, we have defined $q=\exp(2\pi i z_1)$,
$r=\exp(2\pi i z_2)$ and $s=\exp(2\pi i z_3)$.
The constants $(a_1,a_2,b_1,b_2)$ take values $(0,1)$. 
Thus there are sixteen genus-two theta constants. The
even theta constants are those for which
$\mathbf{a}^{\textrm{T}}\mathbf{b}=0\mod 2$. There are ten such theta
constants for which we list the values of $\mathbf{a}$ and $\mathbf{b}$:
\begin{center}
\begin{tabular}{|c|c|c|c|c|c|c|c|c|c|c|}\hline 
$m$ & 0 & 1 & 2 & 3 & 4 & 5 & 6 & 7 & 8 & 9 
\\ \hline 
$\begin{pmatrix} \mathbf{a} \\ \mathbf{b} \end{pmatrix}$ &
$\left(\begin{smallmatrix}0\\0\\0\\0 \end{smallmatrix}\right)$ &
$\left(\begin{smallmatrix}0\\1\\0\\0 \end{smallmatrix}\right)$ &
$\left(\begin{smallmatrix}1\\0\\0\\0 \end{smallmatrix}\right)$ &
$\left(\begin{smallmatrix}1\\1\\0\\0 \end{smallmatrix}\right)$ &
$\left(\begin{smallmatrix}0\\0\\0\\1 \end{smallmatrix}\right)$ &
$\left(\begin{smallmatrix}1\\0\\0\\1 \end{smallmatrix}\right)$ &
$\left(\begin{smallmatrix}0\\0\\1\\0 \end{smallmatrix}\right)$ &
$\left(\begin{smallmatrix}0\\1\\1\\0 \end{smallmatrix}\right)$ &
$\left(\begin{smallmatrix}0\\0\\1\\1 \end{smallmatrix}\right)$ &
$\left(\begin{smallmatrix}1\\1\\1\\1 \end{smallmatrix}\right)$  \\ \hline
\end{tabular}
\end{center}
We will refer to the above ten theta constants as $\theta_m(\mathbf{Z})$
with $m=0,1,\ldots,9$ representing the ten values of 
$\mathbf{a}$ and $\mathbf{b}$ as defined in the above table. Note that six of the even theta constants with $\mathbf{a}\neq0$ have even Fourier coefficients while the remaining four theta constants with $\mathbf{a}=0$ have integral Fourier coefficients.

The modular functions $\Delta_5(\mathbf{Z})$ and 
$\Delta_3(\mathbf{Z})$ can be written out in terms of the even 
theta constants\cite{Nikulin:1995,Raghavan:1976}. One finds
\begin{eqnarray}
\Delta_5(\mathbf{Z}) &=& \frac1{64} \prod_{m=0}^9 \theta_m(\mathbf{Z})\ , \\
\Delta_3(\mathbf{Z}) &=& \frac1{64}\ \theta_2(\mathbf{Z})\! 
\prod_{m=1\textrm{ mod }2} \theta_m(\mathbf{Z})\ .
\end{eqnarray}

Let us define $\widetilde{\Delta}_3(\mathbf{Z})$ to be the square-root of 
$\widetilde{\Phi}_6(\mathbf{Z})$.\begin{equation}
\widetilde{\Delta}_3(\mathbf{Z}) = \frac1{16}\ \theta_1(\mathbf{Z}) \
\theta_3(\mathbf{Z})\ 
\theta_6(\mathbf{Z})\ 
\theta_7(\mathbf{Z})\ 
\theta_8(\mathbf{Z})\ 
\theta_9(\mathbf{Z})\ ,
\end{equation}
squares to given $\tilde{\Phi}_6(\mathbf{Z})$.

Let us denote the square-root of the modular forms,  that appear in the $\BZ_4$ CHL orbifold, $\Phi_{3}(\mathbf{Z})$ and $\widetilde{\Phi}_{3}(\mathbf{Z})$ by  $\Delta_{3/2}(\mathbf{Z})$ and $\widetilde{\Delta}_{3/2}(\mathbf{Z})$ respectively. They can be written in terms of genus-two theta constants. One has
\begin{align}
\Delta_{3/2}(\mathbf{Z})& =\frac18  \theta_5(2\mathbf{Z})\ 
\theta_7(2\mathbf{Z})\ 
\theta_9(2\mathbf{Z})\ ,
 \\
\widetilde{\Delta}_{3/2}(\mathbf{Z}) &=\frac14 \theta_3( \mathbf{Z}')\ 
\theta_8(\mathbf{Z}')\ 
\theta_9(\mathbf{Z}')\ ,
\end{align}
where $\mathbf{Z}'= \left(\begin{matrix} \tfrac12 z_1 & z_2 \\ z_2 & 2 z_3 \end{matrix} \right)$.  Both these modular forms have integral Fourier coefficients as follows from the Fourier coefficients of the theta constants.

\subsection{Notation}
Let $F(\tau,z)$ be a Jacobi form of weight $k$ and index $m$. Under a modular transformation, $\gamma \in SL(2,\mathbb{Z})$, we define
\begin{equation}
F(\tau,z)\big|_\gamma\equiv \exp(-2\pi \imath\tfrac{cmz^2}{c\tau+d})\ (c\tau+d)^{-k}\ F(\gamma\cdot  \tau,\gamma\cdot z )\ ,
\end{equation}
where $\gamma\cdot \tau = \tfrac{a\tau+b}{c\tau+d}$ and $\gamma\cdot z=\tfrac{z}{c\tau+d}$ for $\gamma=\left(\begin{smallmatrix} a & b \\ c & d \end{smallmatrix}\right)$.  This definition is valid even when we are interested in modular forms of congruence subgroups of $SL(2,\mathbb{Z})$. In such cases, those elements of $SL(2,\mathbb{Z})$ that are not in the subgroup of interest will lead to modular forms of some other subgroup that is related to the original subgroup via conjugation. 

\subsection{Eisenstein series at level $N$}

\subsubsection{Prime $N$}

Let $E_2^*(\tau)$ denote the weight two non-holomorphic modular form of $SL(2,\mathbb{Z})$. It is given by
\begin{equation}
E_{2}^*(\tau)=1-24\ \sum_{n=1}^\infty \sigma_{1}(n)\ q^n\   -\frac{3}{\pi\ \textrm{Im} \tau}\ ,
\end{equation}
where $\sigma_{\ell}(n)=\sum_{1\leq d|n} d^\ell$.
The combination\footnote{We caution the reader that the subscript $N$  denotes the level and \textit{not} the weight of the Eisenstein series. All Eisenstein series considered in this paper are of weight two.} 
\begin{equation}
E_{N}(\tau)=\frac1{N-1}\Big(N E_2^*(N\tau)-E_2^*(\tau)\Big)= \tfrac{12i}{\pi(N-1)}\partial_\tau \big[\ln 
\eta(\tau) -\ln \eta(N\tau)\big]
%\frac1{N-1}\Big(N E_2(N\tau)-E_2(\tau)\Big) \ ,
\end{equation}
is a weight two holomorphic modular form of $\Gamma_0(N)$ with constant coefficient equal to $1$\cite[Theorem 5.8]{Stein}. Note the cancellation of the non-holomorphic pieces. Thus, at level $N>1$, the Eisenstein series produces a weight two modular form. For example\footnote{All expansions for the Eisenstein series given here have been obtained using the mathematics software SAGE\cite{sage}. We are grateful to the authors of SAGE for making their software freely available. It was easy for us to verify Eq. \eqref{ourid} using SAGE to the desired order.},
\begin{equation}
E_2(\tau)=1+24 q + 24 q^2 +96 q^3 +24 q^4+144 q^5+96 q^6 +\cdots
\end{equation}
is the weight-two Eisenstein series at level $2$. At levels $3$ and $5$, one has
\begin{eqnarray}
E_3(\tau)  & =&  1 + 12 q +  36 q^2 + 12 q^3 + 84 q^4 +  72 q^5 + 36q^{6}  +\cdots \nonumber \\
     E_5(\tau)  & =&  1 + 6 q +  18 q^2 + 24 q^3 + 42 q^4 +  6 q^5 +72q^{6} 
      + \cdots 
\end{eqnarray}

\subsubsection{Composite $N$}

Suppose $M|N$, then one has $\Gamma_0(N)\subset \Gamma_0(M)$. Thus, for composite $N$, the Eisenstein series at level $M$ is also a modular form at level $N$. For instance at level four, one has two Eisenstein series: $E_2(\tau)$ and
\begin{equation}
 E_4(\tau)  =1 + 8 q + 24 q^2 + 32 q^3 + 24 q^4 + 48 q^5 + \cdots
\end{equation}
At level six, one has three Eisenstein series:  $E_2(\tau)$, $E_3(\tau)$ and
\begin{equation}
 \widehat{E}_6(\tau)  =5/24 + q + 3q^2 + 4q^3 + 7q^4 + 6q^5 +\cdots
\end{equation}
At level eight, one has three Eisenstein series:  $E_2(\tau)$, $E_4(\tau)$ and
\begin{equation}
 \widehat{E}_8(\tau)  =7/24 + q + 3 q^2 + 4 q^3 + 7 q^4 + 6 q^5 + \cdots
\end{equation}
$\widehat{E}_N(\tau)$ refer to Eisenstein series normalized such that the coefficient of $q$ is $+1$. It is known that all Eisenstein series in this normalization have integral coefficients except for the constant term\cite{Stein}.

\subsection{Fourier transform about the cusp at 0}

The modular transformation, $S$, under which $\tau \rightarrow -1/\tau$ maps the cusp at $0$ to the cusp at $i\infty$. When $N$ is prime, $\Gamma_0(N)$ has only these two cusps. One may wish to obtain the Fourier expansion about the cusp at $0$ -- this is done by mapping $0$ to $i\infty$ using the $S$ transform. To obtain the transform of the Eisenstein series, 
first consider
\begin{multline}
E_2^*(N\tau)\big|_S = (\tau)^{-2} \ E_2^*(N S\cdot \tau)\\  
= (\tau)^{-2} \ E_2^*(-N/\tau) = (\tau)^{-2} (\tau/N)^2 E_2^*(\tau/N) = \frac1{N^2} \ E_2^*\left(\tfrac\tau{N}\right)\ .
\end{multline}
Using this result, it is easy to see that
\begin{equation}
E_N(\tau)\big|_S = -\frac1N E_N\left(\tfrac\tau{N}\right)\ .
\end{equation}
Note that $\tau=0$ in the LHS corresponds to $\tau=i\infty$ in the RHS of the above equation. Thus, given the Fourier expansion at $i\infty$, we can obtain the Fourier expansion about $0$. Notice the appearance of fractional powers of $q$, $q^{1/N}$ to be precise, at this cusp. This is expected as the width of the cusp at $0$ is $N$. Also, note that the above formula is valid for all $N$, not necessarily prime. 

Another useful addition formula for the Eisenstein series is the following: 
\begin{equation}\label{ourid}
E_4(\tau) + E_4(\tau+\tfrac12) =2\ E_2(2\tau)\ .
\end{equation}
This formula was experimentally obtained by us and its veracity has been checked to around twenty orders in the Fourier expansion.

\subsection{Fourier transform about other cusps}

The same method can be used to obtain the expansion about other cusps. Again we will need to map the cusp to $i \infty$ and then track the transformation of the non-holomorphic Eisenstein series. Let us do a specific example that is of interest in this paper. Let $N=4$ and consider the cusp at $1/2$.  $\gamma=\left(\begin{smallmatrix} 1 & -1 \\ 2 & -1 \end{smallmatrix}\right)$ maps $1/2$ to $i\infty$.
\begin{multline}
E_4(\tau)\big|_{ST^2S} = -\frac14 E_4(\tfrac\tau4)\big|_{ST^2} = -\frac14 E_4(\tfrac\tau4+\tfrac12)\big|_{S} \\
=-\frac14 \left(2 E_2(\tfrac\tau2)\big|_{S}- E_4(\tfrac\tau4)\big|_{S}\right) 
=\left( E_2(\tau)- E_4(\tau)\right)
\end{multline}
In the penultimate step, we made use of Eq. \eqref{ourid} in order to write $E_4(\tfrac\tau4+\tfrac12)$ in terms of objects with known $S$-transformations.
The final answer is in terms of Eisenstein series whose Fourier coefficients are known thus giving us the expansion of $E_4(\tau)$ about the cusp at $1/2$.

For the CHL models with $N=6$ and $N=8$, it appears that there are no standard methods to determined the Fourier expansion of $E_6(\tau)$ and $E_8(\tau)$ about all the cusps -- this is a minor technical hurdle that needs to be surmounted to complete the computation of the twisted elliptic genus in the corresponding CHL models. It would be helpful if one can obtain identities similar to the one given in Eq. \eqref{ourid}.

\section{The twisted elliptic genus}\label{twistedellgen}

First, let us define the twisted elliptic genus for a $ \mathbb{Z}_N $-orbifold of $K3$:
\begin{equation}
F^{r,s}(\tau,z) = \tfrac1N \textrm{Tr}_{RR,g^r} \Big((-)^{F_L+F_R} g^s q^{L_0}\bar{q}^{\bar{L}_0} e^{2\pi\imath z F_L}\Big)\ ,\quad 0\leq r,s\leq (N-1)\ ,
\end{equation}
where $ g $ generates $ \mathbb{Z}_N $ and $ q=\exp(2\pi \imath \tau) $. We will figure out the $ F^{r,s}(\tau,z) $ by use of their transformation properties under the modular group. We shall do them in several steps. Let $\gamma= \left(\begin{smallmatrix} a & b \\ c & d \end{smallmatrix}\right) \in SL(2,\mathbb{Z})$. Then, one has
\begin{equation}
F^{r,s}(\tau,z)\Big|_\gamma = F^{ar+cs,br+ds}(\tau,z)\ .
\end{equation}
In particular, under $ T:\tau\rightarrow \tau+1 $ and $ S:\tau\rightarrow -1/\tau $, one has
\begin{equation}
F^{0,s}(\tau,z)\Big|_T = F^{0,s}(\tau,z)\quad,\quad
F^{0,s}(\tau,z)\Big|_S = F^{s,0}(\tau,z)\quad.
\end{equation}
More generally, the $ F^{r,s}(\tau,z) $ are weak Jacobi forms of weight zero and index one at level $N$.

\subsubsection*{Step 1: Forming T-orbits}
In step 1, we study the action of $T$ on the $ F^{r,s}(\tau,z) $ and break them up into orbits. 
\begin{itemize}
  \item We have already seen that $F^{0,s}(\tau,z)$ are $T$-invariant i.e., they form orbits of length one.
  \item When gcd$(r,N)=1$, all the $ F^{r,s}(\tau,z) $ form a single orbit of length $N$ (under repeated action of $T$).
  \item When gcd$(r,N)=m$, then the $ F^{r,s}(\tau,z) $ break up into $m$ distinct orbits of length $N/m$. 
\end{itemize}
We will use these results to impose constraints on the form of the $ F^{r,s}(\tau,z) $.

\subsubsection*{Step 2: Ansatz for $F^{0,s}(\tau,z)$}
{\bf Claim:} It suffices to work out $F^{0,s}(\tau,z)$ and the other $ F^{r,s}(\tau,z) $ can be obtained by the action of suitable $SL(2,\mathbb{Z})$ operations. 
 
In step 2, we write out the most general $F^{0,s}(\tau,z)$. Using  proposition 6.1 of\cite{Aoki:2005} for weak Jacobi forms of $ \Gamma_0^J(N)$, $F^{0,s}(\tau,z)$ can be written as follows:
\begin{eqnarray}
F^{0,0}(\tau,z)&=& \tfrac2N A(\tau,z)\quad, \\
F^{0,s}(\tau,z) &=& a\ A(\tau,z) + \alpha_N(\tau)\ B(\tau,z)\ ,\ \ s\neq 0\ ,
\end{eqnarray}
where $ \alpha_N(\tau) $ is a weight-two modular form of $\Gamma_0(N)$ and
\begin{equation}\label{ABdef}
A(z_1,z_2)
=  \sum_{i=2}^4 \left(\frac{\vartheta_i(z_1,z_2)}{\vartheta_i(z_1,0)}\right)^2 
\quad,\quad
B(z_1,z_2)=
\left(\frac{\vartheta_1(z_1,z_2)}{\eta^3(z_1)}\right)^2 \ .
%\label{weakJacobiForm}
\end{equation}

When $N$ is composite, the  dimension of modular forms at weight two is greater than one. We list the possibilities for $ N=4,6,8 $.
\begin{eqnarray}
\alpha_4(\tau)&=& b_1\ E_2(\tau) +b_2\ E_4(\tau) \ , \\ 
\alpha_6(\tau)&=& b_1\ E_2(\tau) +b_2\ E_3(\tau) + b_3\ E_6(\tau)\ , \\
\alpha_8(\tau)&=& b_1\ E_2(\tau) +b_2\ E_4(\tau) + b_3\ E_8(\tau)\ ,
\end{eqnarray}
where $ E_N(\tau) $ is the Eisenstein series of weight-two and level $N$:
$$
E_N(\tau)=\tfrac{12i}{\pi(N-1)}\partial_\tau \big[\ln \eta(\tau) -\ln \eta(N\tau)\big]\ ,
$$
normalized so that its constant coefficient is one.

\subsubsection*{Step 3: Imposing constraints from sizes of T-orbits}
In step 3, we study the $S$ transformation on our ansatz for $F^{0,s}(\tau,z)$ and then follow its transformation under powers of $T$ and make the ansatz for $\alpha_N(\tau)$ compatible with its orbit size. 
\begin{itemize}
  \item When $(s,N)=1$, there are no obvious constraints.
  \item When  $(s,N)=m>1$, then there will be constraints.
  \begin{itemize}
    \item When $N=4$ and $s=2$, then $b_2=0$ as we need to have an orbit of size two.
    \item When $N=6$ and $s=2,4$, then $b_1=b_3=0$ so that it is consistent with an orbit size of three. 
    \item When $N=6$ and $s=3$, then $b_2=b_3=0$ so that it is consistent with an orbit size of two. 
    \item When $N=8$ and $s=2,6$, then $b_3=0$ so that it is consistent with an orbit size of four.
    \item When $N=8$ and $s=4$, then $b_2=b_3=0$ so that it is consistent with an orbit size of two.
  \end{itemize}
\end{itemize}
Further simplification occurs when we consider the symmetry, $ F^{r,s}(\tau,z)=F^{-r,-s}(\tau,z) $. It implies that we have the equivalence $ F^{0,s}(\tau,z)=F^{0,N-s}(\tau,z) $. 
\begin{itemize}
  \item For $ N=4 $, we need to only work out $ F^{0,0}(\tau,z) $, $ F^{0,1}(\tau,z) $ and $ F^{0,2}(\tau,z) $.
  \item For $ N=6 $, we need to only work out $ F^{0,0}(\tau,z) $, $ F^{0,1}(\tau,z) $, $ F^{0,2}(\tau,z) $ and $ F^{0,3}(\tau,z) $.
  \item For $ N=8 $, we need to only work out $ F^{0,0}(\tau,z) $, $ F^{0,1}(\tau,z) $, $ F^{0,2}(\tau,z) $, $ F^{0,3}(\tau,z) $ and $ F^{0,4}(\tau,z) $.
\end{itemize}

\subsubsection*{Step 4: Using topological data}
In the next step, we fix the undetermined constants by studying the conditions on the Fourier coefficients, $ c_b^{0,s}(-1) $ and $ c_b^{0,s}(0) $ of $F^{0,s}(\tau,z)$. As shown by David, Jatkar and Sen\cite{David:2006ru}, these two sets of numbers are related to topological objects on $K3$ and hence can be determined by studying the action of the group on $ H^*(K3,\mathbb{Z}) $. Let $Q^{0,s}$ be the number of $g^s$-invariant elements of $ H^*(K3,\mathbb{Z}) $ (where $g$ generates $\mathbb{Z}_N$). David, Jatkar and Sen show that (see Eq. \eqref{fourierdefs} for the definition of $c_m^{a,b}$)
\begin{equation}
Q^{0,s} = N c_0^{0,s}(0) + 2N c_1^{0,s}(-1)\ .
\end{equation}
Further $N c_1^{0,s}(-1)$ counts the number of $g^s$-invariant $(0,0)$ and $(0,2)$ forms on $K3$. For symplectic involutions, these forms are invariant and hence $N c_1^{0,s}(-1)=2$. (When $N=11$, the involution is  non-symplectic and here we expect the answer to be different.) We thus obtain the relation
\begin{equation}
N c_0^{0,s}(0) = Q^{0,s}-4\ .
\end{equation}
 
 It is easy to compute $Q^{0,s}$ given the cycle shape which we shall do now.
\begin{itemize}
  \item {\bf Prime} $N$: The cycle shape is $1^{k+2}N^{k+2}$. When, $s=0$, all forms contribute and hence $Q^{0,0}=24$. For any $s\neq0$, one has $Q^{0,s}=k+2$. This implies that $ N c_0^{0,0}(0)=20 $ and $ N c_0^{0,s}(0)=k-2 $ for $s\neq0$.
  \item $N=4$: The cycle shape is $1^4 2^2 4^4$. This implies that $Q^{0,1}=Q^{0,3}=4$ and $Q^{0,2}=8$. We thus obtain $ 4 c_0^{0,s}(0) =0$ for $s=1,3$ while $ 4 c_0^{0,2}(0) =4$.
  \item $N=6$: The cycle shape is $1^2 2^2 3^26^2$. This implies that $Q^{0,1}=Q^{0,5}=2$ and $Q^{0,2}=Q^{0,4}=6$ and $ Q^{0,3}=8 $. Thus one has $ 6 c_0^{0,s}(0) =-2$ for $s=1,5$, $ 6 c_0^{0,3}(0) =4$  and $ 6 c_0^{0,s}(0) =2$ for $s=2,4$.
  \item $N=8$: The cycle shape is $1^2 2^1 4^18^2$. This implies that $Q^{0,1}=Q^{0,3}=Q^{0,5}=Q^{0,7}=2$ and $Q^{0,2}=Q^{0,6}=4$ and $ Q^{0,4}=8 $. Thus one has $ 8 c_0^{0,s}(0) =-2$ for $s=1,3,5,7$, $ 8 c_0^{0,s}(0)=0$ for $s=2,6$ and $ 8 c_0^{0,4}(0) =4$.
\end{itemize}
Further, one has
\begin{equation}
\label{eqn:type}
c_0^{0,0}(0)=\tfrac{20}N \quad,\quad  c_1^{0,s}(-1)=\tfrac2N \ .
\end{equation}
A nice consistency check is to verify that $k=\tfrac12 \sum_{s=0}^{N-1} c_0^{0,s}(0)$. This relation holds in all our examples.

\subsubsection*{Step 5: Fixing undetermined parameters}

Steps 1-4 are the same for all $N$, whether prime or composite. For prime $N$, at the end of step 4, no undetermined parameters remain. However, for composite $N$, this 
is not true. For $N=4$, there is one undetermined parameter in $F^{0,1}(\tau,z)$. For $N=6$, there are two undetermined parameters and for $N=8$, there are five undetermined parameters. These will have to be dealt with on a case by case basis and we will illustrate the procedure for $N=4$ in this paper. The occurrence of additional cusps for composite $N$ is the key to fixing these parameters.

\section{Explicit Formulae}

Below we provide the initial terms in the Fourier expansion of the modular forms $\Delta_k(\mathbf{Z})$ defined in this paper
\begin{eqnarray*}
 \Delta_5&=&\left( - \frac{1}{{\sqrt{r}}} 
         + {\sqrt{r}} \right) \,\sqrt{q}{\sqrt{s}} + 
 \left( \frac{9}{r^{\frac{5}{2}}} - 
     \frac{93}{r^{\frac{3}{2}}} + 
     \frac{90}{{\sqrt{r}}} - 90\,{\sqrt{r}} + 
     93\,r^{\frac{3}{2}} - 9\,r^{\frac{5}{2}} \right) \,
   q^{\frac{3}{2}} s^{\frac{3}{2}} \\
   &+& \left( r^{- \frac{3}{2} } + 
     \frac{9}{{\sqrt{r}}} - 9\,{\sqrt{r}} - 
     r^{\frac{3}{2}} \right) \,
   \left( q^{\frac{3}{2}}\,{\sqrt{s}} + 
     {\sqrt{q}}\,s^{\frac{3}{2}} \right)  \\ 
     &+ &
  \left( \frac{-9}{r^{\frac{3}{2}}} - 
     \frac{27}{{\sqrt{r}}} + 27\,{\sqrt{r}} + 
     9\,r^{\frac{3}{2}} \right) \,
   \left( q^{\frac{5}{2}}\,{\sqrt{s}} + 
     {\sqrt{q}}\,s^{\frac{5}{2}} \right)  \\
     &+& 
  \left( -r^{- \frac{5}{2}  } + 
     \frac{27}{r^{\frac{3}{2}}} + 
     \frac{12}{{\sqrt{r}}} - 12\,{\sqrt{r}} - 
     27\,r^{\frac{3}{2}} + r^{\frac{5}{2}} \right) \,
   \left( q^{\frac{7}{2}}\,{\sqrt{s}} + 
     {\sqrt{q}}\,s^{\frac{7}{2}} \right) \\ 
     &+& 
  \left( \frac{9}{r^{\frac{5}{2}}} - 
     \frac{12}{r^{\frac{3}{2}}} + 
     \frac{90}{{\sqrt{r}}} - 90\,{\sqrt{r}} + 
     12\,r^{\frac{3}{2}} - 9\,r^{\frac{5}{2}} \right) \,
   \left( q^{\frac{9}{2}}\,{\sqrt{s}} + 
     {\sqrt{q}}\,s^{\frac{9}{2}} \right) \\ 
     &+ &
  \left( \frac{-27}{r^{\frac{5}{2}}} - 
     \frac{90}{r^{\frac{3}{2}}} - 
     \frac{135}{{\sqrt{r}}} + 135\,{\sqrt{r}} + 
     90\,r^{\frac{3}{2}} + 27\,r^{\frac{5}{2}} \right) \,
   \left( q^{\frac{11}{2}}\,{\sqrt{s}} + 
     {\sqrt{q}}\,s^{\frac{11}{2}} \right)  \\
     &+&
  \left( r^{- \frac{7}{2}  } + 
     \frac{12}{r^{\frac{5}{2}}} + 
     \frac{135}{r^{\frac{3}{2}}} - 
     \frac{54}{{\sqrt{r}}} + 54\,{\sqrt{r}} - 
     135\,r^{\frac{3}{2}} - 12\,r^{\frac{5}{2}} - 
     r^{\frac{7}{2}} \right) \,
   \left( q^{\frac{13}{2}}\,{\sqrt{s}} + 
     {\sqrt{q}}\,s^{\frac{13}{2}} \right)  +\cdots
%      \\+ 
%   \left( \frac{-9}{r^{\frac{7}{2}}} + 
%      \frac{90}{r^{\frac{5}{2}}} + 
%      \frac{54}{r^{\frac{3}{2}}} + 
%      \frac{99}{{\sqrt{r}}} - 99\,{\sqrt{r}} - 
%      54\,r^{\tfrac{3}{2}} - 90\,r^{\tfrac{5}{2}} + 
%      9\,r^{\tfrac{7}{2}} \right) \,
%    \left( q^{\tfrac{15}{2}}\,s^{\tfrac12} + 
%      q^{\tfrac{5}{2}}\,s^{\tfrac{3}{2}} + 
%      q^{\tfrac{3}{2}}\,s^{\tfrac{5}{2}} + 
%      q^{\tfrac12}\,s^{\tfrac{15}{2}} \right)
\end{eqnarray*}
\begin{eqnarray*}
\Delta_{3/2}&=&\left(\sqrt{r}-\frac{1}{\sqrt{r}}\right)
   \sqrt{s}
   \sqrt{q}
   +\left(r^{5/2}-r^{3/2}+2
   \sqrt{r}-\frac{2}{\sqrt{r}}+\frac{1}{r^{3/2}}-\frac{1}{
   r^{5/2}}\right) s^{3/2}   q^{3/2} \\
   && +\left(-r^{3/2}+\sqrt{r}-\frac{1}{\sqrt{r
   }}+\frac{1}{r^{3/2}}\right)
   \left(\sqrt{q} s^{3/2}+\sqrt{s} q^{3/2}\right) \\
   &&+\left(-r^{3/2}+\sqrt{r}-\frac{1}{\sqrt{r}
   }+\frac{1}{r^{3/2}}\right)
  \left(\sqrt{q}s^{5/2}+ \sqrt{s}q^{5/2}\right)
   \\
   &&+\left(r^{5/2}-r^{3/2}+2
   \sqrt{r}-\frac{2}{\sqrt{r}}+\frac{1}{r^{3/2}}-\frac{1}{
   r^{5/2}}\right) \left(\sqrt{q} s^{7/2}+\sqrt{s} q^{7/2}\right)
 \\ && +\left(r^{5/2}-2
   r^{3/2}+\frac{2}{r^{3/2}}-\frac{1}{r^{5/2}}\right)
  \left(\sqrt{q} s^{9/2}+ \sqrt{s} q^{9/2}\right)
   \\
   &&+\left(r^{5/2}+\sqrt{r}-\frac{1}{\sqrt{r}}
   -\frac{1}{r^{5/2}}\right)
   \left(\sqrt{q} s^{11/2}+\sqrt{s} q^{11/2}\right)
  \\
   && +\left(-r^{7/2}+2 r^{5/2}-r^{3/2}+2
   \sqrt{r}-\frac{2}{\sqrt{r}}+\frac{1}{r^{3/2}}-\frac{2}{
   r^{5/2}}+\frac{1}{r^{7/2}}\right)
   \left(\sqrt{q} s^{13/2}+\sqrt{s} q^{13/2}\right)+\cdots
\end{eqnarray*}
\begin{eqnarray*}
\widetilde{\Delta}_{3/2}&=& \left( -\frac{1}{{\sqrt{r}}}   + 
        {\sqrt{r}} \right) \,{\sqrt{s}} 
   {\sqrt{q_h}}+\left(2r^{3/2}-2
   \sqrt{r}+\frac{2}{\sqrt{r}}-\frac{2}{r^{3/2}}\right) s^{3/2}   {q_h}^{3/2} \\
   &&+ 
  \left( \frac{2}{{\sqrt{r}}} - 2\,{\sqrt{r}}
        \right) \,
        \left(\sqrt{q_h} {s}^{\frac{3}{2}}+\sqrt{s} {q_h}^{\frac{3}{2}}\right) + 
   \left( \frac{2}{{\sqrt{r}}} - 2\,{\sqrt{r}}
        \right)\left(\sqrt{q_h} {s}^{\frac{5}{2}}+\sqrt{s} {q_h}^{\frac{5}{2}} \right) 
   \\
   && + 
  \left( \frac{-4}{{\sqrt{r}}} + 
        4\,{\sqrt{r}} \right)  \left(\sqrt{q_h} {s}^{\frac{7}{2}}+\sqrt{s} {q_h}^{\frac{7}{2}}  \right) \\
       && + 
  \left( r^{- \frac{3}{2}  } - 
        \frac{2}{{\sqrt{r}}} + 2\,{\sqrt{r}} - 
        r^{\frac{3}{2}} \right)  \left(\sqrt{q_h}{s}^{\frac{9}{2}} +\sqrt{s}{q_h}^{\frac{9}{2}}  \right) \\
  &&  + 
   \left( \frac{-2}{r^{\frac{3}{2}}} + 
        2\,r^{\frac{3}{2}} \right) \left(\sqrt{q_h}{s}^{\frac{11}{2}}+\sqrt{s}{q_h}^{\frac{11}{2}}  \right) \, \\
   && + 
  \left( \frac{-2}{r^{\frac{3}{2}}} + 
        \frac{4}{{\sqrt{r}}} - 4\,{\sqrt{r}} + 
        2\,r^{\frac{3}{2}} \right) \left( \sqrt{q_h} {s}^{\frac{13}{2}} +\sqrt{s} {q_h}^{\frac{13}{2}}  \right)+\cdots \ ,
\end{eqnarray*}
where $q_h\equiv q^{1/4}$. The expression is symmetric under the exchange $q\leftrightarrow s^4$ and antisymmetric under $r\rightarrow r^{-1}$. An all-orders proof follows from the properties of the even genus-two theta constants.

\subsection{Invariance of $\widetilde{\Phi}_3(\mathbf{Z})$}\label{lastone}

Under the level $4$ subgroup, $G_0(4)$, of $Sp(2,\mathbb{Z})$, the modular form $\widetilde{\Phi}_3(\mathbf{Z})$ transforms as
\begin{equation}\label{phi3transform}
\widetilde{\Phi}_3(M\cdot \mathbf{Z}) = \left(\frac{-1}{\det D}\right)\ \det(C\mathbf{Z}+D)^3\ \widetilde{\Phi}_3(\mathbf{Z})\ ,
\end{equation}
where $M=\begin{pmatrix}A & B \\ C & D \end{pmatrix}\in Sp(2,\BZ)$, $M\cdot \mathbf{Z}= (A \mathbf{Z}+ B)(C \mathbf{Z}+D)^{-1}$ and $C=0 \mod 4$. 

Consider the subgroup of $G_0(4)$ given by $B=C=0$ and $A^{\textrm{T}}=D^{-1}$. Under this subgroup, 
Eq. \eqref{phi3transform} can be written as
\begin{equation}
\widetilde{\Phi}_3(D^{\textrm{T}}\cdot \mathbf{Z}\cdot  D) =  \left(\frac{-1}{\det D}\right)\ (\det D)^3\  \widetilde{\Phi}_3(\mathbf{Z})\ .
\end{equation}
Choosing $D=\gamma=\begin{pmatrix} 1 & -1 \\ 4 & -3 \end{pmatrix}$, one sees that $\widetilde{\Phi}_3(\mathbf{Z})$ is invariant since the Jacobi symbol as well as $\det D=+1$. Similarly,  when $D=\delta=\begin{pmatrix} -1 & 1 \\ 0 & 1 \end{pmatrix}$, again the Jacobi symbol as well as $\det D=-1$ leading to $\widetilde{\Phi}_3(\mathbf{Z})$ being invariant under the $G_0(4)$ transformation generated by $\delta$.
\bibliography{master}
\end{document}